\documentclass[conference]{IEEEtran}
\IEEEoverridecommandlockouts
\usepackage{cite}
\usepackage{amsmath,amssymb,amsfonts}
\usepackage{algorithmic}
\usepackage{graphicx}
\usepackage{textcomp}
\usepackage{graphicx}
\usepackage{dcolumn}
\usepackage{bm}
\usepackage{hyperref}
\usepackage{tikz}
\usepackage{mathtools}
\usepackage{physics}
\usepackage[normalem]{ulem}
\usepackage{braket}
\usepackage{hyperref}
\usepackage{multirow}
\usepackage{stackengine}
\usepackage{array}
\usepackage{xcolor}
\usepackage{lipsum}
\usepackage{enumitem}
\usepackage{dblfloatfix} 
\usepackage[caption=false]{subfig}
\usepackage{lipsum}
\usepackage{array}

\definecolor{cincinnati-red}{RGB}{190,0,0}
\definecolor{ForestGreen}{RGB}{34,139,34}

\def\BibTeX{{\rm B\kern-.05em{\sc i\kern-.025em b}\kern-.08em
    T\kern-.1667em\lower.7ex\hbox{E}\kern-.125emX}}
\begin{document}

\title{Distance-Independent Entanglement Generation in a Quantum Network using Space-Time Multiplexed Greenberger–Horne–Zeilinger (GHZ) Measurements\\
\thanks{This work was funded by the National Science Foundation (NSF) grant  CNS-1955834 and NSF ERC Center for Quantum Networks (CQN) grant EEC-1941583.  \\Accepted in IEEE QCE 2021.\copyright2021 IEEE}
}

\author{\IEEEauthorblockN{Ashlesha Patil $^{\dagger}$}
\IEEEauthorblockA{\textit{Wyant College of Optical Sciences} \\
\textit{University of Arizona, Tucson AZ USA}\\
ashlesha@email.arizona.edu}
\and
\IEEEauthorblockN{Joshua I. Jacobson $^{\dagger}$}
\IEEEauthorblockA{\textit{Department of Physics,} \\
\textit{University of Arizona, Tucson AZ USA}\\
jijacob1@email.arizona.edu}
\and
\IEEEauthorblockN{Emily Van Milligen}
\IEEEauthorblockA{\textit{Department of Physics,}\\
\textit{University of Arizona, Tucson AZ USA}\\
evanmilligen@email.arizona.edu}
\and
\IEEEauthorblockN{Don Towsley}
\IEEEauthorblockA{\textit{College of Information and Computer Sciences,} \\
\textit{University of Massachusetts, Amherst MA USA}\\
towsley@cs.umass.edu}
\and
\IEEEauthorblockN{Saikat Guha}
\IEEEauthorblockA{\textit{Wyant College of Optical Sciences} \\
\textit{University of Arizona, Tucson AZ USA}\\
saikat@optics.arizona.edu}

}

\maketitle

\begin{abstract}
In a quantum network that successfully creates \textit{links}---shared Bell states between neighboring repeater nodes---with probability $p$ in each time slot, and performs Bell State Measurements at nodes with success probability $q<1$, the end-to-end entanglement generation rate drops exponentially with the distance between consumers, despite multi-path routing. If repeaters can perform multi-qubit projective measurements in the GHZ basis that succeed with probability $q$, the rate does not change with distance in a certain $(p,q)$ region, but decays exponentially outside. This region where the distance-independent rate occurs is the super-critical region of a new \textit{percolation} problem. We extend this GHZ protocol to incorporate a time-multiplexing blocklength $k$, the number of time slots over which a repeater can mix-and-match successful links to perform fusion on. As $k$ increases, the super-critical region expands. For a given $(p,q)$, the entanglement rate initially increases with $k$, and once inside the super-critical region for a high enough $k$, it decays as $1/k$ GHZ states per time slot. When memory coherence time exponentially distributed with mean $\mu$ is incorporated, it is seen that increasing $k$ does not indefinitely increase the super-critical region; it has a hard $\mu$-dependent limit. Finally, we find that incorporating space-division multiplexing, i.e., running the above protocol independently in up to $d$ disconnected network regions, where $d$ is the network’s node degree, one can go beyond the 1 GHZ state per time slot rate that the above randomized local-link-state protocol cannot surpass. As $(p,q)$ increases, one can approach the ultimate min-cut entanglement-generation capacity of $d$ GHZ states per slot.
\end{abstract}

\def\thefootnote{$\dagger$}\footnotetext{These authors contributed equally to this work.}

\section{Introduction}
\label{sec:intro}

Quantum networks are designed to distribute shared entangled states, which serve as an important resource for applications such as quantum key distribution (QKD), entanglement enhanced quantum sensing, and distributed quantum computing. Quantum networks are composed of fundamental building blocks called {\em quantum repeaters}, which are special purpose quantum computers that can store qubits (entangled, e.g., with qubits at neighboring repeaters) in a local quantum memory register, and perform quantum logic and measurements on sets of locally-held qubits. It is well known that, without the assistance of quantum repeaters, the shared entanglement generation rate drops exponentially with the distance between communicating parties~\cite{pirandola2017fundamental,Takeoka2014-bh}. Several attempts have been made to improve the rate-versus-distance scaling starting with ideal as well as noisy Bell states generated across network links, followed by measurements performed at repeater nodes using local link state information, aimed at lengthening the reach of entanglement as well as to purify entanglement across links or multiple hops. In this paper, we will be concerned with {\em quantum switches}, but will continue to call them a repeater. A quantum switch is a repeater that has the capability of choosing which links to fuse, based on instantaneous network state information, dynamically. 

With ideal two-qubit maximally-entangled Bell states generated with probability $p$ across each network in each time slot and repeaters allowed to perform Bell State Measurements (BSMs) that succeed with probability $q$, \cite{pant2019routing} showed that the path diversity (multi-path routing) afforded by the network improves the exponent of the exponential rate-distance scaling over what is achievable when the network routes entanglement over a pre-determined path. However, the rate still decays exponentially even with global link state knowledge as long as $q < 1$. Ref.~\cite{dhara2021sub} showed that time multiplexing over a linear chain of repeaters, i.e., allowing a repeater node to wait for blocks of $k$ time slots before deciding the pairs of qubits (within that time block) to be attempted to be fused (using BSMs), with an optimally-chosen $k$ can achieve a sub-exponential entanglement rate-vs.-distance scaling. 

For the case when link-level Bell states have sub-unity fidelity, ref.~\cite{Yuan2020quantum} designed a quantum repeater architecture that uses multiplexing and a photonic switchboard to maximize the tradeoff between the shared entanglement rate and the fidelity of final entangled state. Ref.~\cite{Tim} on the other hand optimized the length of time that quantum memories should hold onto noisy Bell states to maximize the rate. Ref.~\cite{victora2020purification} performed an extensive numerical optimization of the scheduling of purification and BSMs to maximize the distillable entanglement rate, given a target fidelity for the final shared entanglement state.

Following the intuition of ~\cite{dhara2021sub}, ref.~\cite{Emily} evaluates the benefits of time-multiplexing in a two-dimensional quantum network. This paper extends the multi-path entanglement routing protocol presented in ref.~\cite{pant2019routing} by allowing quantum memories to hold onto entangled qubits for multiple time-steps before deciding which qubit pairs to perform BSMs on. In the absence of quantum memory decoherence, time multiplexing can only help bridge the gap between the rate obtained using local link state knowledge and that obtained using global link state knowledge. When the BSM success probability $q<1$, as the multiplexing block length increases, the obtainable rate monotonically increases. When memory decoherence is accounted for, the rate increases until some optimal time-multiplexing block length is reached, and then proceeds to decrease. In both cases there are regions of the parameter space ($p$, $q$, and multiplexing block length $k$) where this dynamic local-link state knowledge based routing protocol can outperform simply routing over the predetermined shortest paths between end-consumers.

All of the above papers achieve rates that decrease with the distance between the communicating parties. In \cite{patil2020entanglement} however, it was shown that if repeaters are allowed to perform GHZ joint projective measurements (on $3$ or more qubits), then in a certain $(p,q)$ region---determined by the critical threshold of a modified site-bond percolation problem---the entanglement rate does not decay with the end-to-end distance. This paper aims to improve the maximum entanglement of the protocol described in ~\cite{patil2020entanglement} by employing time-multiplexed GHZ projections, as well as spatial-division multiplexing. Our intent is to quantify the regime of the parameter space, i.e., $p$, the probability that a network edge establishes a (noiseless) Bell state across it successfully in each time slot, $q$, the probability that an attempted GHZ measurement succeeds, and $k$, the time-multiplexing block length, which affords distance-independent entanglement rate. As the network has to wait for $k$ time steps before the entangled state is shared between the consumers, the entanglement rate is defined as no. of shared entangled states generated per time slot and not ebits/s. We employ two techniques to realize these improvements: (a) time-multiplexing to attain distance-independent rate for poorer $(p,q)$ parameters, and (b) spatial-division multiplexing to increase the maximum achievable rate. We also study the performance of the time-multiplexed protocol for the case that quantum memories are non-ideal and have finite coherence times that are exponentially-distributed with a known mean. This paper serves as a guidebook to determine the degree of time-multiplexing needed to achieve distance independent rate, but not use too high a $k$ once inside the percolation region so as not to be faced with the $1/k$ rate penalty measured in end-to-end entangled states generated per time slot, \& decide if spatial multiplexing is needed, given the knowledge of $p$ and $q$.

Finally, all mention of `qubit' in this paper should be interpreted as being logical (error-corrected) qubits, and not physical qubits. This is because loss on a logical qubit can be {\em heralded} with the error-detection and error-correction property of a quantum code. This is an important assumption in modeling assumptions inherent in this paper. More discussion on this aspect will appear in Section~\ref{sec:decoherence}.

The organization of the paper is as follows. We start with a discussion of the time-multiplexed version of the GHZ measurement based protocol in Section~\ref{sub:time_mux_protocol} followed by the percolation problem the protocol translates to in Section~\ref{sub:percolation}. The simulation results for the protocol are analysed in Section~\ref{sub:time_mux_results}. We relax the ideal quantum memory assumption made in Section~\ref{sub:time_mux_protocol} to discuss how the finite memory coherence times affect the rates and the optimal degree time-multiplexing in Section~\ref{sec:decoherence}. We discuss the spatially multiplexed version of the protocol in Section~\ref{sec:net_div}, where we subdivide the network into spatially-disconnected regions while employing our protocol independently within each region, to achieve entanglement rate that approaches the ultimate capacity, the min-cut rate~\cite{pirandola2019end}, for the underlying network topology. We conclude by summarising the main results in Section~\ref{sec:conclusion}. The code used to generate the simulation results is available at \cite{code}.

\section{Time-multiplexed GHZ-projection based entanglement routing}
\label{sec:time_mux}

In this section, we introduce our time-multiplexed protocol where nodes perform up to $n$-qubit GHZ measurements, with qubits chosen over $k$ time slots. We term this the $(n,k)$-GHZ protocol. We describe the protocol, present simulated entanglement rates for a square grid network, followed by a new percolation problem our protocol maps to. 

Let us begin with a few definitions. An $n$-qubit maximally entangled state of the form $\frac{\ket{0}^{\otimes n}+\ket{1}^{\otimes n}}{\sqrt{2}}$ is referred to as the $n$-GHZ state, which includes the Bell state $\frac{\ket{00}+\ket{11}}{\sqrt{2}}$, as a special case ($n=2$). There are $2^n$ mutually-orthogonal $n$-GHZ states, forming an orthonormal basis for the Hilbert space spanned by $n$ qubits. Two fundamental quantum operations that are used in our entanglement-routing protocol, are as follows: 

\noindent (1) \textit{$n$-fusion}---This is a probabilistic operation, which when successful, implements a joint projective measurement on $n$ qubits (held at a repeater node) in the aforesaid $n$-GHZ basis. We assume that the $n$-fusion measurement projects the $n$ measured qubits on to one of the $2^n$ GHZ states, $\frac{\ket{0}^{\otimes n}+\ket{1}^{\otimes n}}{\sqrt{2}}$. This does not result in a loss of generality, since if the $n$-fusion happens to project the $n$ qubits on to any of the other $2^{n}-1$ $n$-GHZ states, one can apply local single-qubit Pauli unitary operations to bring the post-measurement state to the one obtained when the fusion projects on to the aforesaid standard $n$-GHZ state. A $2$-fusion is same as the BSM. An $n$-fusion can be used to fuse $n$ entangled states, e.g., cluster states, each with ${m_1,m_2,\dots, m_n}$ qubits respectively, into one entangled state with $m=\sum_{i=0}^n m_i-n$ qubits. If the constituent entangled states are each GHZ states, the resulting fused state is an $m$-GHZ state. The failure of an $n$-fusion attempt (on $n$ qubits within a memory register at a given repeater) is taken to mean that all of those $n$ qubits are measured in the Pauli-X basis independently, and the (binary) results of those measurements are known to that repeater, and are communicated to Alice and Bob at the end of each $k$-time-slot block.

\noindent (2) \textit{Pauli-X basis measurement}---When performed on one qubit of an $n$-GHZ state, an $X$ basis measurement removes that qubit from the GHZ state, resulting in an $n-1$ GHZ state. This operation is used to trim the shared entangled state such that only the communicating parties (the `consumers') have qubits left over that are (potentially) entangled at the end of each round of the protocol. Pauli-X basis measurements are also used to model fusion failures as mentioned above. 

\subsection{The time-multiplexed entanglement-routing protocol}
\label{sub:time_mux_protocol}
We consider a quantum network with repeaters placed at nodes. The network topology can be a regular lattice such as a square-grid or hexagonal, or a random network of a known node-degree distribution, such as the configuration graph~\cite{Newman2001-fz}. We refer to a shared Bell pair, $\frac{\ket{00}+\ket{11}}{\sqrt{2}}$ between two neighboring repeater nodes as a \textit{link}. In each time slot, of length $\tau$ seconds, an attempt is made to create a link across every network edge; it succeeds with probability $p$. The two qubits of a link are stored in quantum memory registers at the two nodes sharing the edge, if successful. The knowledge of the success-failure of each link takes time $\tau_L = L/c$ seconds, from the time when the entanglement attempt is initiated, to arrive at the nodes sharing that edge. Here, $L$ (meters) is the physical distance of the network edge, and $c$ is the speed of light. 

We define the {\em time-multiplexing block length}, $k$, as the number of time-steps a repeater waits---in accumulating qubits entangled with neighboring nodes---before attempting fusions on a subset of the qubits accumulated, whose success-failure outcomes have arrived at the node. Fusions are destructive measurements. Performing an $n$-fusion thus frees up $n$ qubit slots in the memory register at the node. It should therefore be evident that at any given point in time, each repeater node holds between $d\left(\lceil{\tau_L/\tau}\rceil\right)$ and $d\left(\lceil{\tau_L/\tau}\rceil + k\right)$ qubits, where $d$ is the node's degree ($d=4$ for a square-grid network).

\begin{figure}
    \centering
    \includegraphics[scale = 0.55]{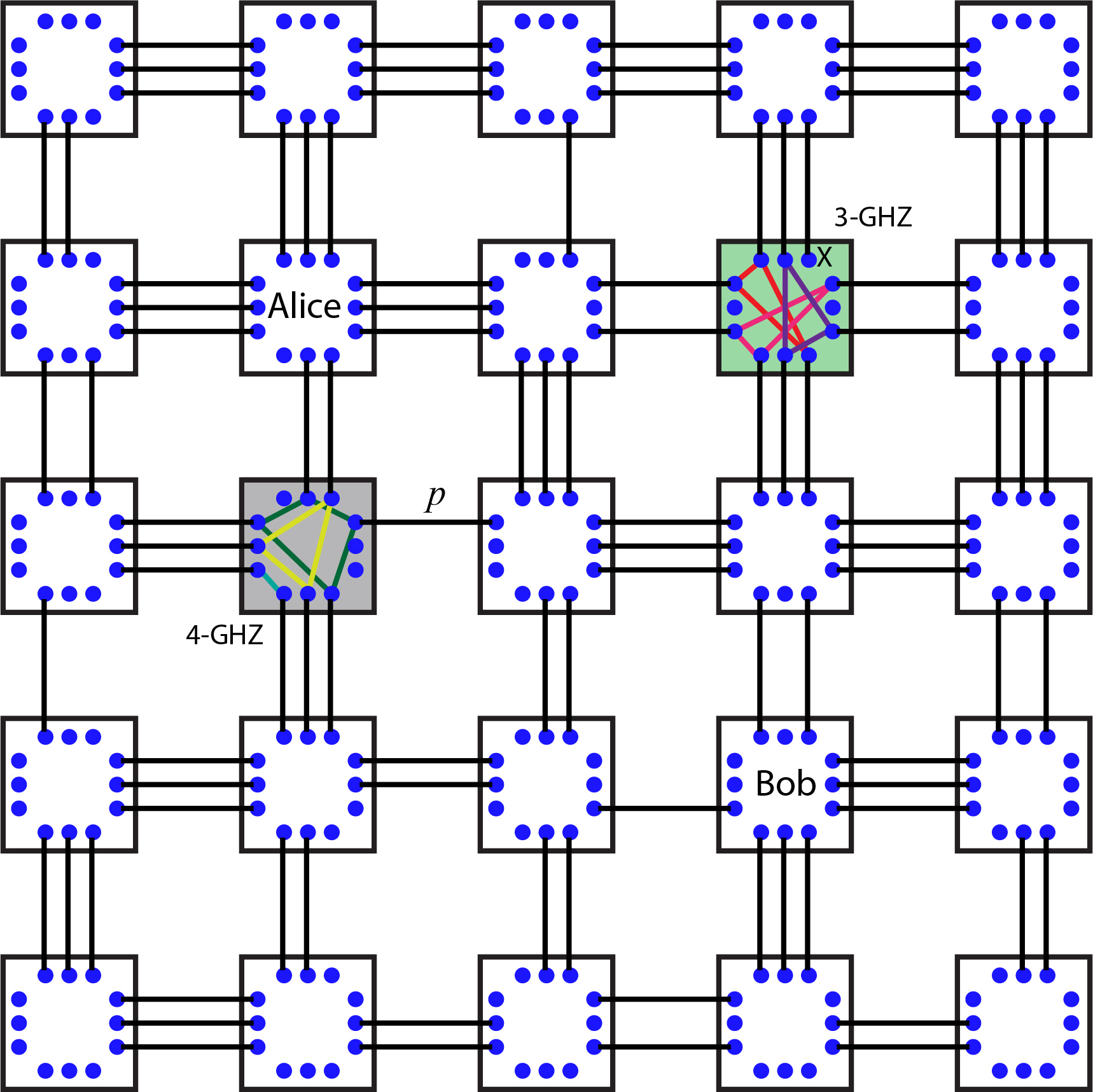}
    \caption{Schematic depiction of a random snapshot of our $(n,k)$-GHZ protocol with $k=3$ over a square-grid quantum network. The black squares are quantum repeater nodes and blue dots are qubits held within an ideal quantum memory register within the nodes. Entanglement links (two-qubit Bell pairs) are shown using black lines. In each time slot, a link is successfully established across each network edge with probability $p$. The lines connecting qubits within a node illustrate fusion groups. Fusions are attempted within each node, based on local link-state information after each subsequent $k$-time-slot block. Each fusion attempt succeeds with probability $q$. The maximum number of qubits that can comprise a fusion group is $n$. The repeaters shaded green and grey show instances of fusion-group formations within a node, for the $(3,3)$-GHZ and $(4,3)$-GHZ protocols, respectively.}
    \label{fig:time_mux_schm}
\end{figure}

Every repeater can perform one- and two-qubit unitary operations and Pauli basis measurements on qubits held in its memory register. Fig.~\ref{fig:time_mux_schm} depicts a $k=3$ link-snapshot over a square-grid network. The line segments denote successful link attempts over a $3$ time slot block. The blue dots represent qubits. In this section, we assume that the quantum memory registers are ideal in the sense that the qubit coherence time is much larger than ${\tau_L} + (k-1)\tau$ seconds.

In our $(n,k)$-GHZ protocol, depicted schematically in Fig.~\ref{fig:time_mux_schm}, at the end of each block of $k$ time slots, nodes other than Alice or Bob, the communicating parties (the `consumers'), perform fusions on groups of (up to $n$) qubits, to expand the reach of the shared entanglement across the network. The rules for picking the fusion groups are described below. Each repeater acts based on local link state knowledge, i.e., the link success-failure outcomes on its neighboring edges. Fusion outcomes are independent events, each with probability of success $q$. 

Let us use $d$ to denote node degree. For a square-grid network, $d = 4$. We will evaluate the performance of the $(n,k)$-GHZ protocol described below, on the square-grid network, for $n=3$ and $n=4$.
\begin{itemize}
    \item A repeater randomly picks one successful link across neighboring edges, over one $k$ time slot block, into groups of (at least $2$ and at most $\min(d,n)$) qubits to attempt fusions on. It repeats the process until there are no more links across two (or more) distinct neighbors. All links within the $k$-slot block are treated as equal in forming fusion groups, i.e., qubits of a link do not get preferential treatment based on the time-slot they were created in.
    \item Each group of qubits to fuse must be entangled with qubits across distinct neighboring edges.
    \item A repeater maximizes the number of $n$-fusions that can be performed with the available links, followed by maximizing $(n-1)$-fusions, and so on.
    \item Qubits that are not part of any fusion group at a node are removed by performing Pauli-X basis measurements.
    \item The measurement results of each of the single-qubit Pauli-X measurements and the multi-qubit fusion measurements are communicated to Alice and Bob at the end of each time slot.
\end{itemize} 

All fusion measurements and Pauli-X measurements at repeater nodes commute with one another. Hence they can be performed without coordination, yet synchronously, in each time slot, based on the protocol described above that uses local link-state information. Because all the qubits across the network in any $k$-time-slot block are measured (and hence destroyed), either with fusion or Pauli-X measurements, at the end of each block of $k$ time slots, Alice and Bob end up with potentially one or more shared GHZ states involving qubits that are held at Alice's and Bob's local memory registers. We define the entanglement generation {\em rate} as the number of GHZ states shared divided by $k$, expressed in the unit of {\em GHZ states per time slot}. Note that when $k=1$, i.e., the $(n,1)$-GHZ protocol reduces to the $n$-GHZ protocol as presented in ~\cite{patil2020entanglement}.

\subsection{Mapping our protocol to a mixed-percolation problem}
\label{sub:percolation}

Let us consider our time-multiplexed $(n,k)$-GHZ protocol on a quantum network with an underlying topology given by a graph $G = (V,E)$ defined over a vertex set $V$ and edge set $E$. The link success probability per time slot $p$, and the success probability of each $n$-fusion $q$, translate respectively to effective bond and site occupation probabilities of a site-bond (mixed) percolation problem. The number of shared GHZ states after the measurements are performed at the end of each $k$-time-slot block equals the number of disjoint subgraphs shared between qubits held by Alice and Bob, in a graph whose nodes are individual qubits and whose edges are entangled links across network edges and edges drawn among qubits of fusion groups within repeater nodes as depicted in Fig.~\ref{fig:time_mux_schm}~\cite{patil2020entanglement}. In the appendix, we present a graphical construction, which shows that employing time multiplexing (i.e., $k > 1$) simply changes the lattice on which this mixed-percolation problem is defined on. The percolation rules remain the same as those for the non-time multiplexed protocol described in~\cite{patil2020entanglement}.

\begin{figure}[htb]
    \centering
    \includegraphics[scale=0.65]{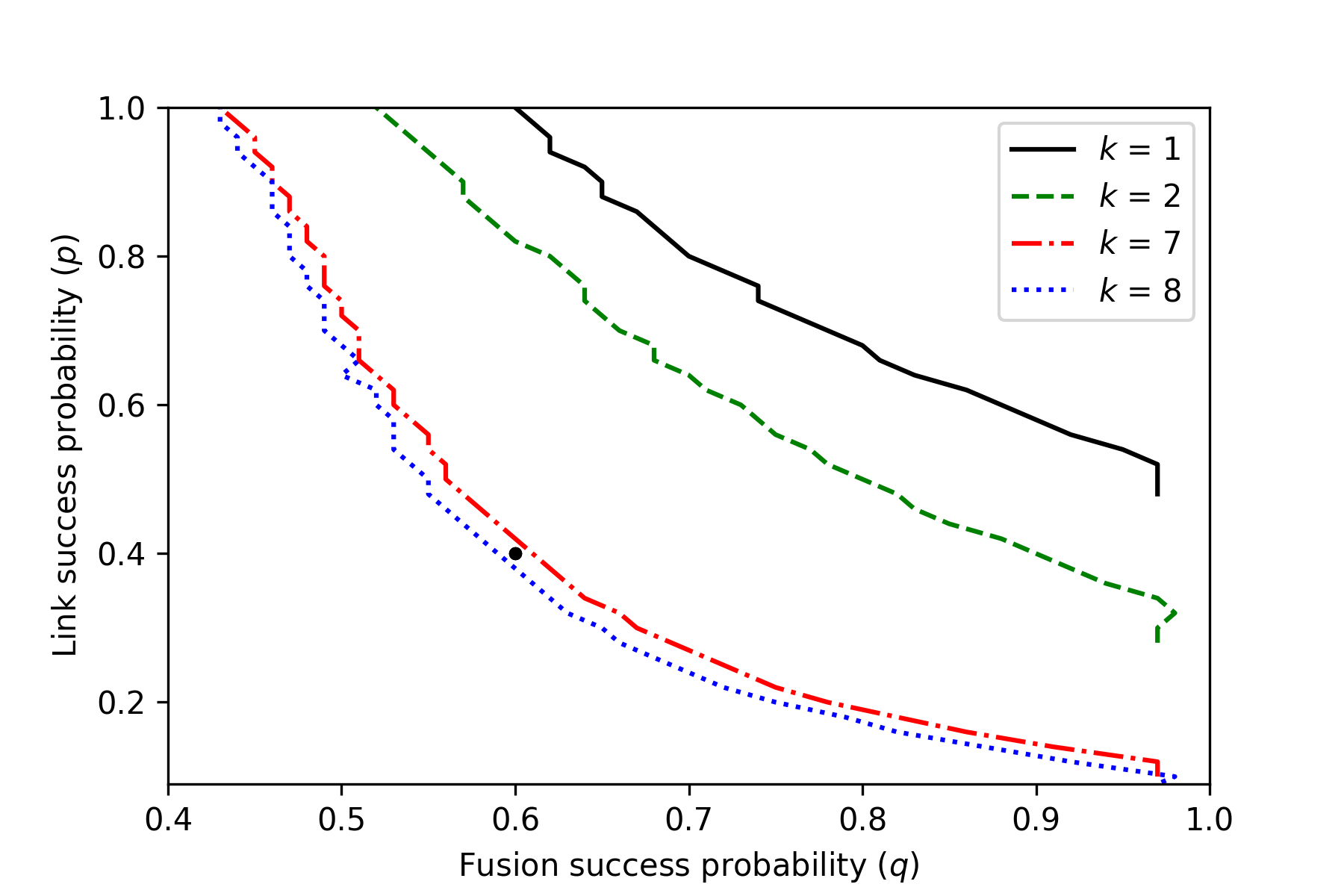}
    \caption{The critical boundaries of the site-bond percolation problem induced by our $(4,k)$-GHZ protocol, on the square-grid network, for $k=1,2,7$ and $8$. The curves are not sufficiently smooth due to finite-size effects of our simulations. To generate each simulated point shown, we average over $50$ runs of a Newman-Ziff~\cite{Newman2001-vm} algorithm variant over a $100$-by-$100$ square-grid.}
    \label{fig:site_bond_4ghz}
\end{figure}

In Fig.~\ref{fig:site_bond_4ghz} we plot the critical boundaries for the site-bond percolation induced by the ($4,k$)-GHZ protocol with increasing time-multiplexing block length $k$. The entanglement generation rate does not scale with distance in the {\em super-critical} region, i.e., above the critical boundary. The entanglement rate decays exponentially with the distance between Alice and Bob on the other side of the critical boundary. Increasing $k$ expands the super-critical region, and hence affords distance-independent rates with progressively lower $(p,q)$ values than is possible with the $k=1$ protocol in~\cite{patil2020entanglement}. Increasing $k$ indefinitely would in principle indefinitely expand the critical boundary outwards. However, after $k=7$ or so, for the square-grid topology, we notice diminishing returns of increasing $k$ further. Further, the maximum $k$ that can be supported would depend upon the maximum quantum memory coherence times. Recall that a coherence time of $\lceil{\tau_L/\tau}\rceil + (k-1)$ times slots (or this times $\tau$ seconds) is necessary to support our protocol.

We observed that our described method to choose fusion groups at repeater nodes results in a larger super-critical region compared to a protocol that randomly groups links at a node without trying to maximize the number of $n$-fusions. However, we also believe that an improved version of our protocol is possible that time-orders link successes akin to~\cite{dhara2021sub}.

\subsection{Simulation results}
\label{sub:time_mux_results}
\begin{figure*}[htb]
\centering
\subfloat[\label{sfig:3GHZ_compare_rVSp_k_d_80}]{%
  \includegraphics[scale = 0.6]{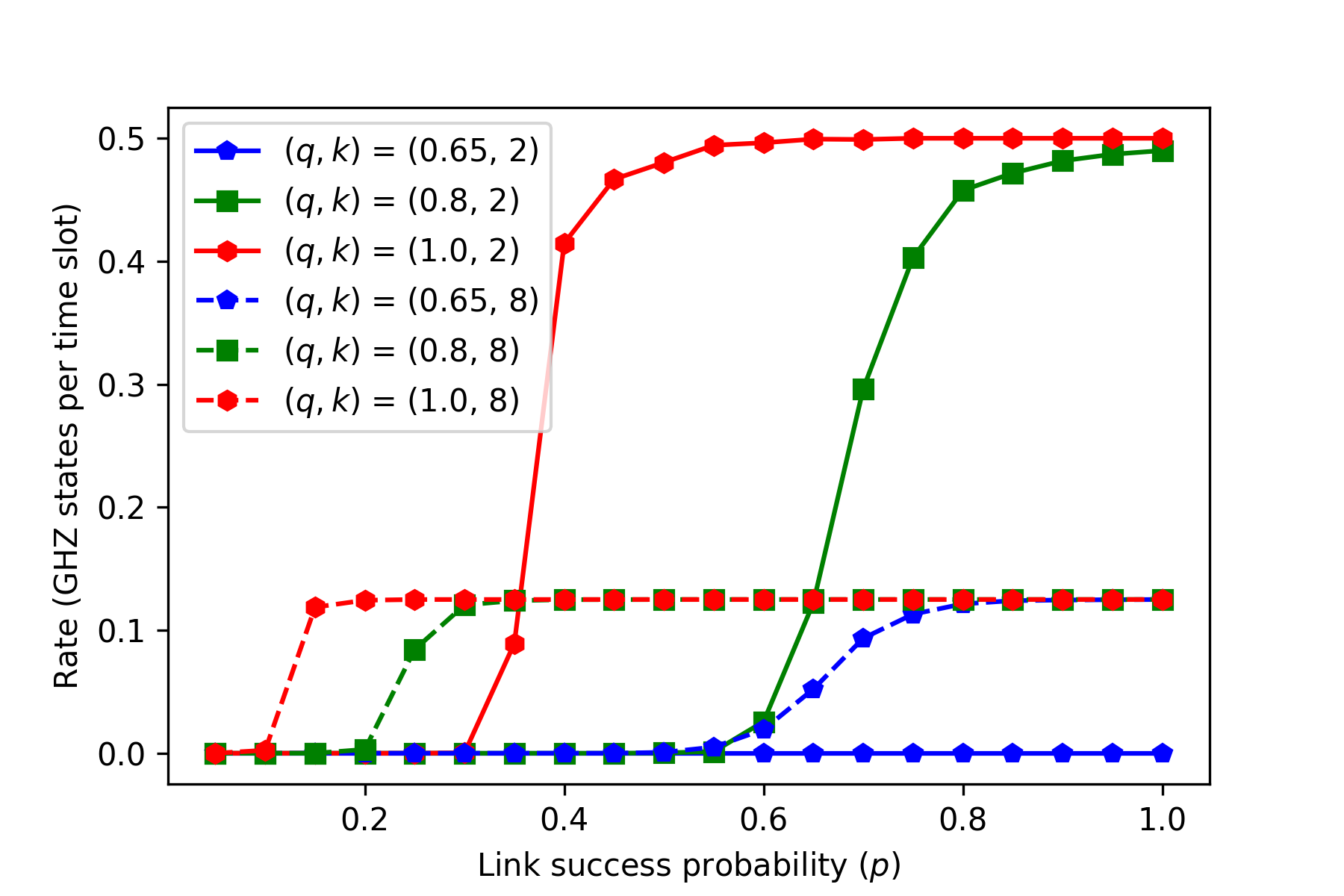}%
}
\subfloat[\label{sfig:3GHZ_compare_rVSq_k_d_80}]{%
  \includegraphics[scale = 0.6]{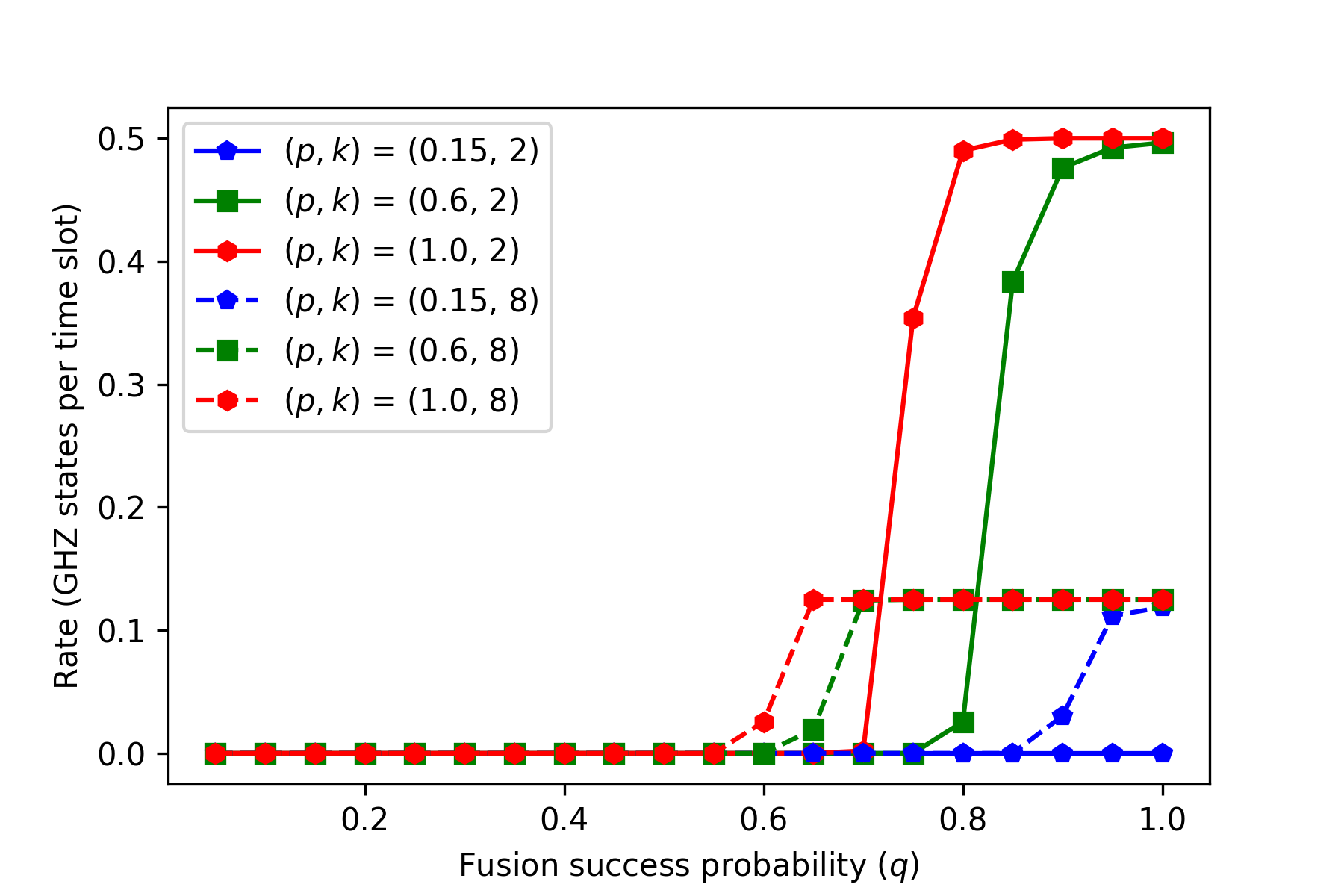}%
}
\vspace{-0.1cm}
\centering
\subfloat[\label{sfig:4GHZ_compare_rVSp_k_d_80}]{%
  \includegraphics[scale = 0.6]{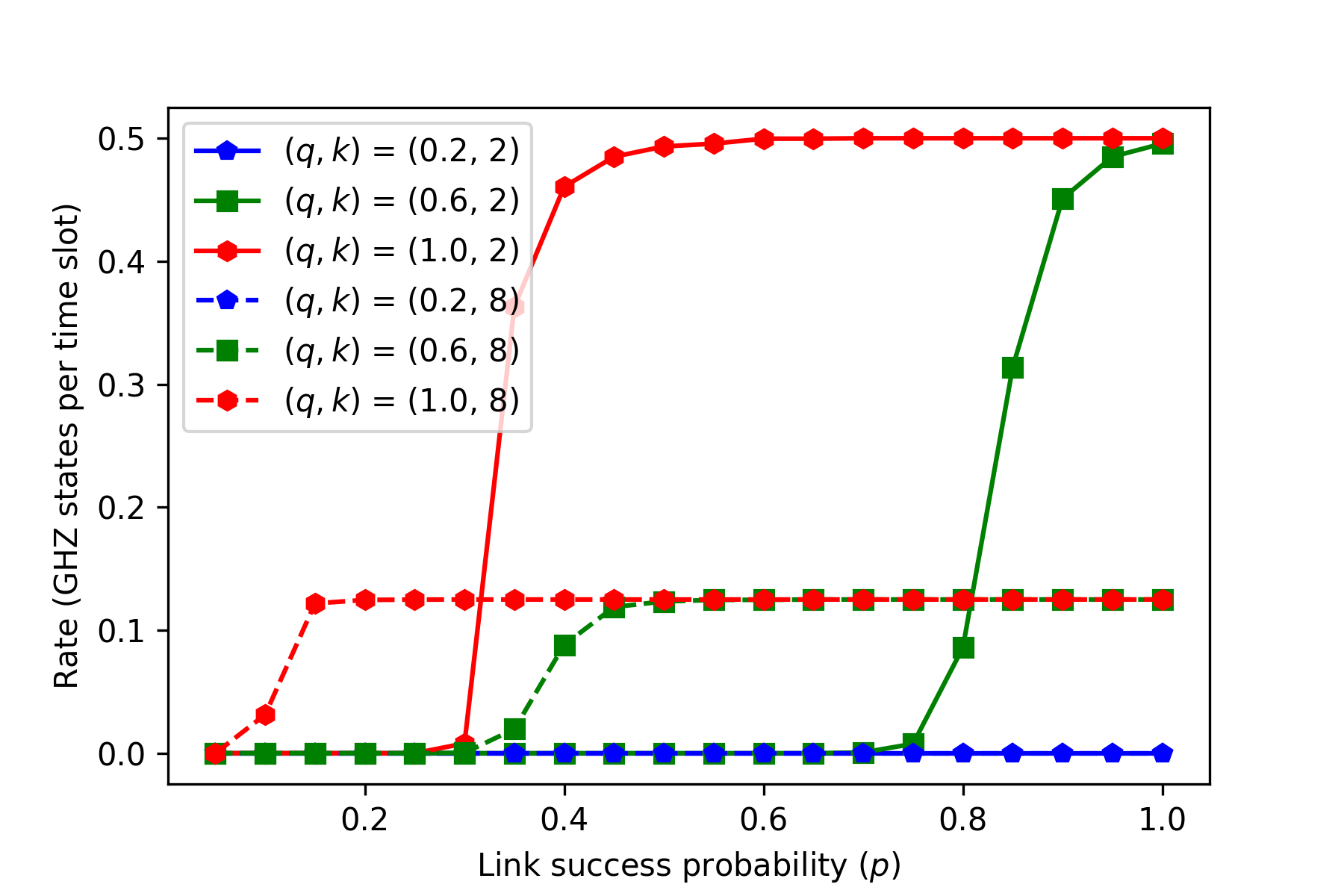}%
}
\subfloat[\label{sfig:4GHZ_compare_rVSq_k_d_80}]{%
  \includegraphics[scale = 0.6]{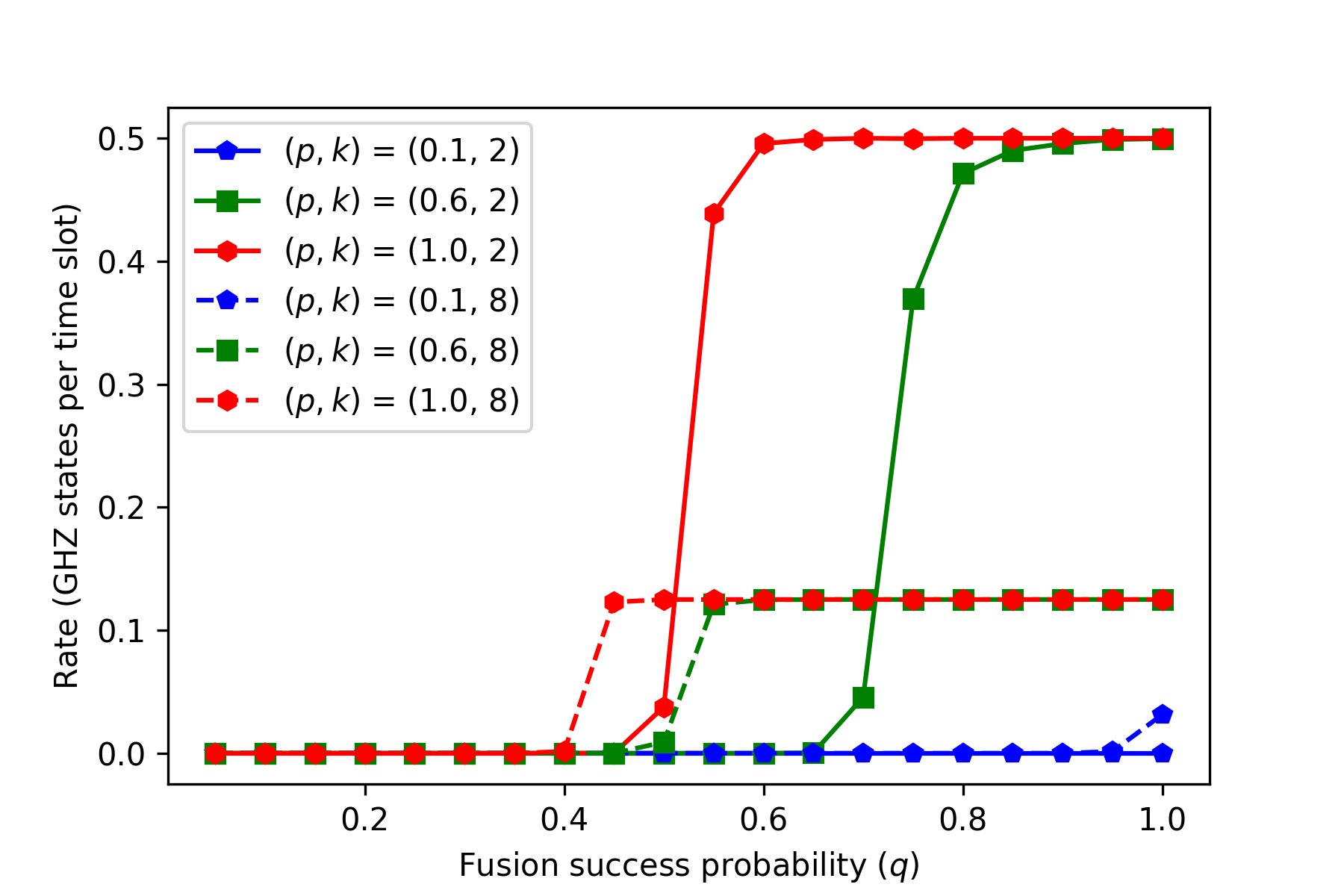}%
}
\caption{Effect of time-multiplexing on entanglement generation rate (in units of shared GHZ states between Alice and Bob per time slot). (a) Rate vs. $p$ for $(3,k)$-GHZ protocol, (b) Rate vs. $q$ for $(3,k)$-GHZ protocol, (c) Rate vs. $p$ for $(4,k)$-GHZ protocol, (b) Rate vs. $q$ for $(4,k)$-GHZ protocol. For each of the four plots, we have chosen the cases of $k=2$ and $k=8$. Solid lines correspond to $k=2$ and dashed lines are for $k=8$. The rate always saturates to $1/k$ GHZ states per time slot. This is because, if the percolation condition is met (i.e., in the super-critical regime), there exists with high probability a unique single giant-connected (spanning) component, which translates to Alice and Bob sharing one entangled GHZ state among qubits held in their memory registers.}
\label{fig:rate_pqk}
\end{figure*}

 All simulation results reported are for the ($4,k)$-GHZ protocol over the square-grid network, unless specified otherwise. We observed qualitatively similar trends for the ($3,k$)-GHZ protocol, but with admittedly smaller percolation regions than their ($4,k$) counterparts. In Fig.~\ref{fig:rate_pqk}, we plot entanglement rate (in units of shared GHZ states between Alice and Bob, per time slot) as a function of link success probability $p$ (fusion success probability $q$) for different values of fusion success probability (link success probability), both for the ($3,k$)- and $(4,k$)-GHZ protocols, for the cases of $k=2$ and $k=8$. As expected from the percolation regions shown in Section~\ref{sub:percolation}, when $k$ goes from 2 to 8 the values of $p$ and $q$ where a sudden jump occurs in the rate decrease. This sharp transition in the rate at certain values of $p$ and $q$, and then saturating to a constant is an essence of percolation. In the flat region of the rate plots, the rate does not change with the distance between Alice and Bob. Below the percolation transitions, the rate diminishes exponentially with distance. The $(p,q)$ pair at which the rate plots transition to a constant lie on the site-bond critical curve for that $k$ value in Fig.~\ref{fig:site_bond_4ghz}. If we keep increasing $(p,q)$ while keeping $k$ constant, the rate always saturates at $1/k$ GHZ states per time slot. This is because increasing $(p,q)$ beyond the transition point pushes the network deeper into the super-critical region, where there exists with high probability one unique giant connected (spanning) component shared between the qubits of Alice and Bob. As a result, in this region, at most one GHZ state is shared between the consumers at the end of each $k$-time-slot block, resulting in the rate of $1/k$ GHZ states per time slot. We observe that time multiplexing gets rid of the `turn around' in the rate that was observed for the 3-GHZ protocol in~\cite{patil2020entanglement}. The turn around occurs in the $(3,1)$-GHZ protocol on the square grid network in~\cite{patil2020entanglement} because if at least one of the two neighboring repeaters has four successful links, the repeater(s) with four links may not choose the link shared with the other repeater for fusion while the other repeater uses the shared link for fusion. This random selection rule ends up effectively disconnecting the two repeaters. However, for the $(3,k)$-GHZ protocol, when $k>2$, the repeaters can perform multiple fusions. As a result, the link which was left unused in the $(3,1)$-GHZ protocol now gets used to in some other fusion with a high probability, keeping the network connected.

\begin{figure}[htb]
    \centering
    \includegraphics[scale = 0.6]{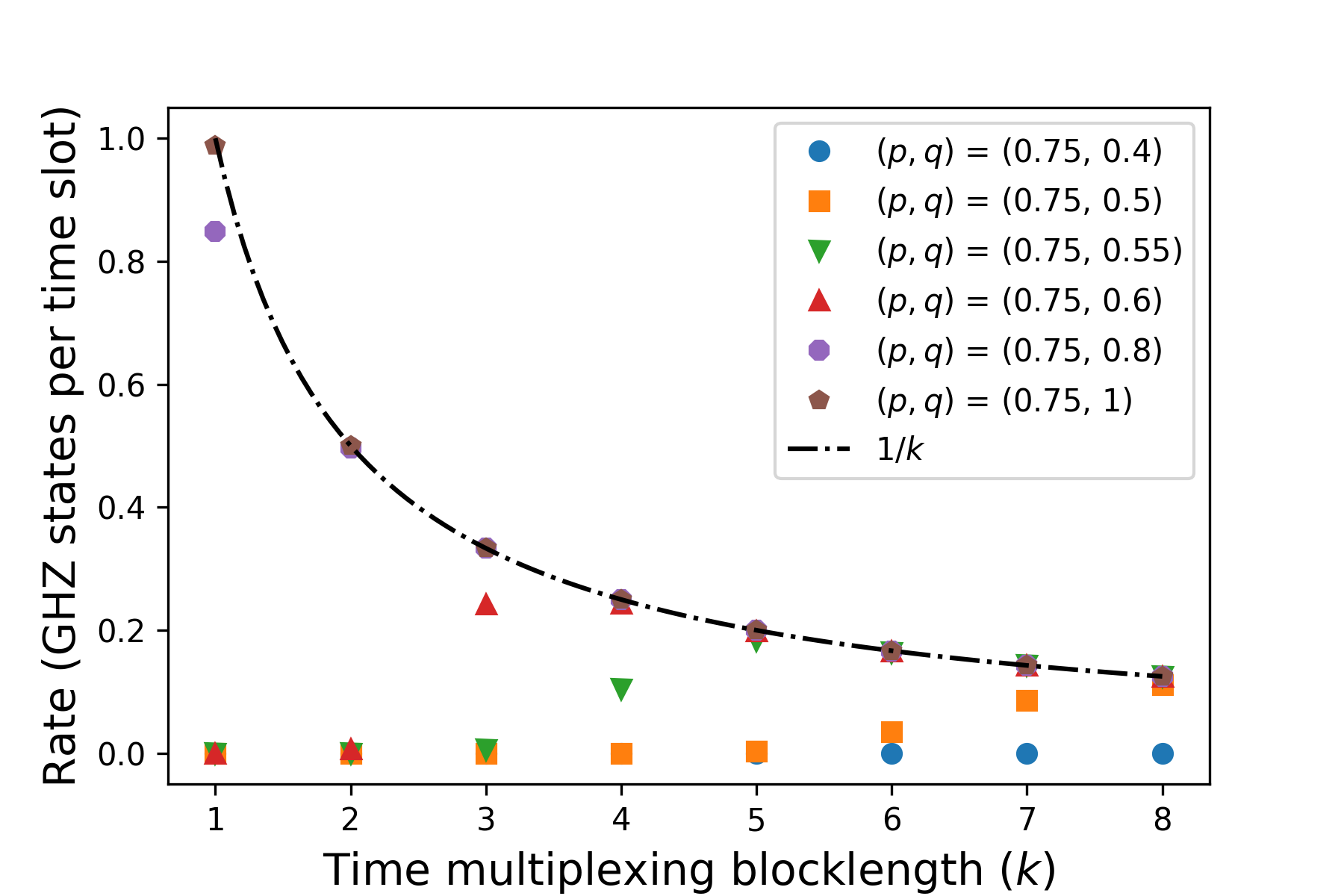}
    \caption{Rate as a function of time multiplexing block-length $k$ for different $(p,q)$ values. The rate increases with increasing $k$ until a threshold $k_{(p,q)}$ beyond which the percolation condition is met, and the rate becomes $1/k$, and  distance independent. }
    \label{fig:4GHZ_plot_rVSk_d_80}
\end{figure}

We keep $(p,q)$ constant and vary $k$ in Fig.~\ref{fig:4GHZ_plot_rVSk_d_80} to demonstrate that time-multiplexing helps when values of $p$ and $q$ are small. We observe that for a given $(p,q)$ pair, the rate increases with $k$ until a certain value of $k$, which we call $k_{(p,q)}$ for that $(p,q)$ pair, at which point the rate hits $1/k_{(p,q)}$ GHZ states per time slot. This is the maximum rate allowed for by the underlying network topology for that $(p,q)$ pair with the $(n,k)$-GHZ protocol. This behaviour can be explained using the percolation plots in Fig.~\ref{fig:site_bond_4ghz}. If we keep $(p,q)$ constant and start increasing $k$, the network barely percolates at $(p,q)$ when $k =k_{(p,q)}$, i.e., $(p,q)$ falls within the super-critical region for the first time when $k =k_{(p,q)}$. If $k$ is increased beyond $k_{(p,q)}$, the rate starts decaying as $1/k$ since $(p,q)$ remains in the super-critical region for larger values of $k$ and the number of shared GHZ states stays $1$. 

If a $(p,q)$ pair falls in between the critical boundaries for two consecutive $k$ values $k_1$ and $k_2=k_1+1$, then the optimal time-multiplexing block length $k_{(p,q)} = k_2$. For example, $(p,q) = (0.4,0.6)$ falls between the critical boundaries for $k=7$ and $k=8$ as seen in Fig.~\ref{fig:site_bond_4ghz}. This $(p,q)$ pair would give exponentially decaying rate with distance for $k=7$ and distance independent rate for $k\geq 8$. The optimal value of time-multiplexing block length therefore, $k_{(0.4,0.6)} = 8$. If $k$ is increased above $8$, the rate will equal $1/k$, but will remain distance-independent. The optimum $k_{(p,q)}$ depends on both the network topology and the $n$ in the $(n,k)$-GHZ protocol, the largest permissible fusion-group size.

\begin{figure*}[hbt]
\centering
\hspace{-8em}
\subfloat[\label{sfig:4GHZ_surf_env_r_p_q_d_80}]{%
  \includegraphics[scale = 0.65]{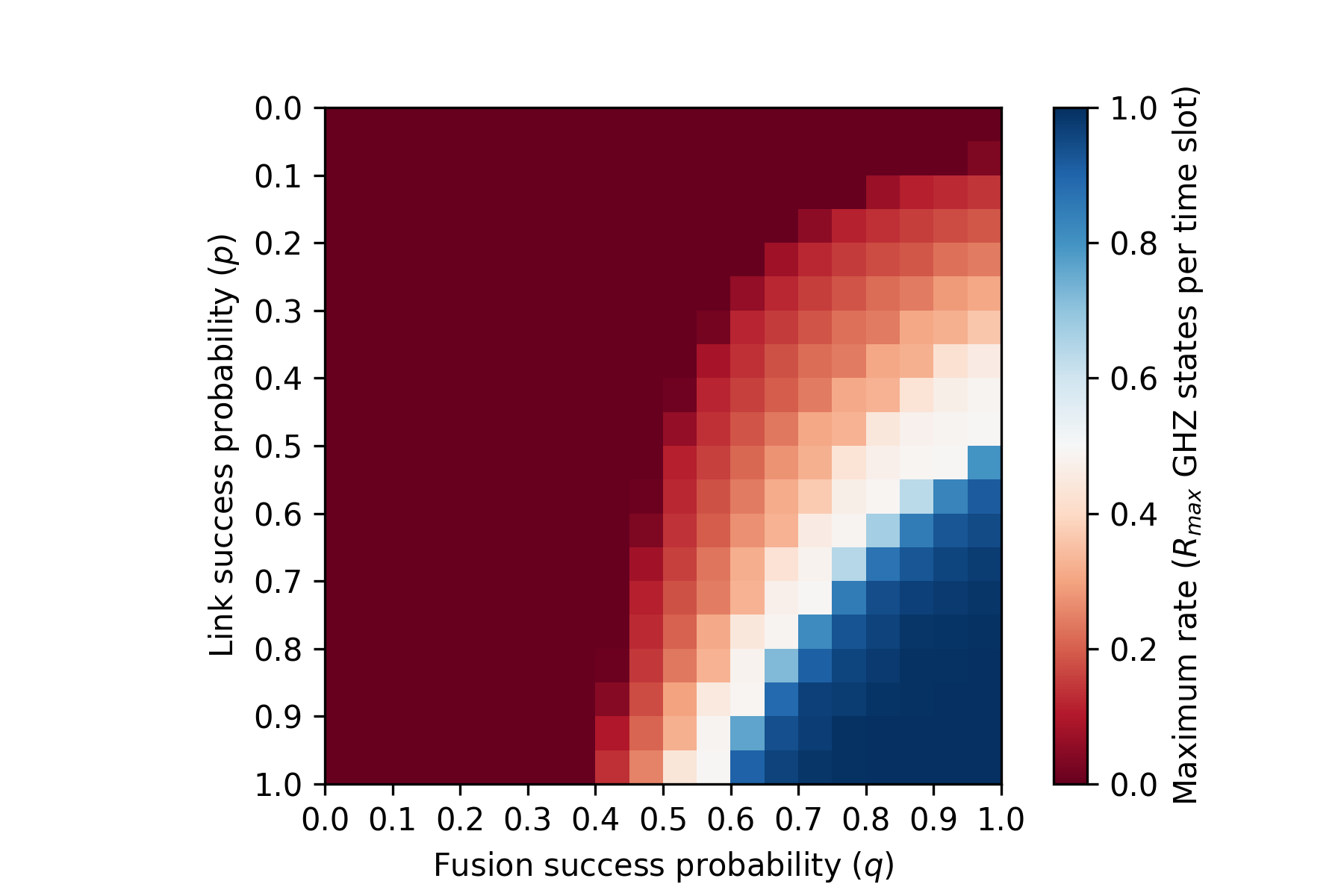}%
}
\subfloat[\label{sfig:4GHZ_surf_env_k_p_q_d_80}]{%
  \includegraphics[scale = 0.65]{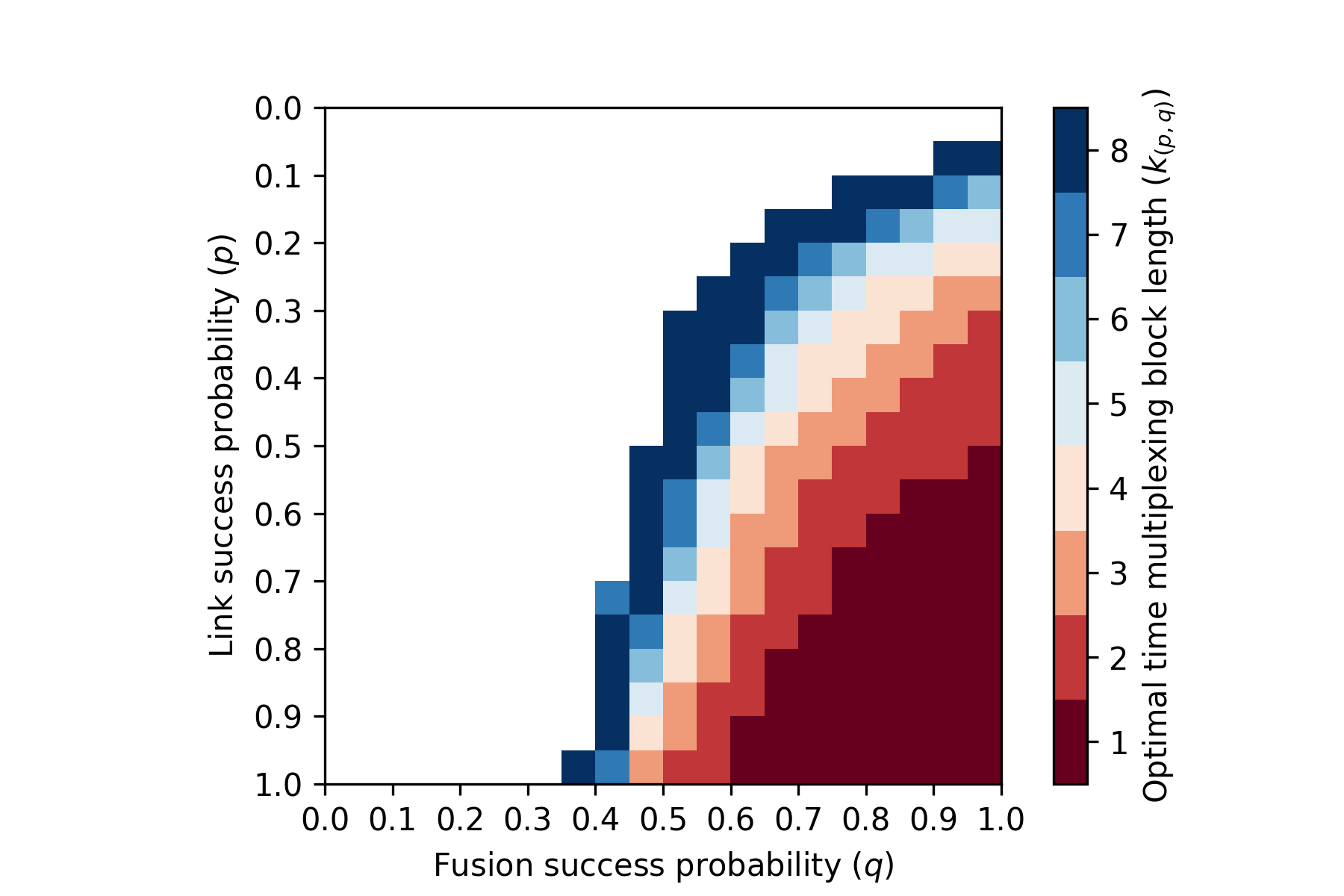}%
}
\caption{(a) Entanglement rate $R_{\max}$ achieved for every $(p,q)$ point maximized over the choice of $k$, for the $(4,k)$-GHZ protocol over a square-grid network; (b) The optimal time multiplexing block length $k_{(p,q)}$ plotted as a heat-map as a function of $p$ and $q$, which attains $R_{\max}$ shown in (a) for every $(p,q)$ value. For both plots, our simulations stop at $k=8$. Hence, $k_{(p,q)}$ does not exist in the white region of (b)}
\label{fig:surf_plots}
\end{figure*}
Fig.~\ref{fig:surf_plots}(a) presents a heat-map of the entanglement rate optimized over the choice of $k$. The blue region with rate very close to $1$ GHZ state per time slot is the super-critical region for $k=1$, the non-time multiplexed protocol. We did simulations only up to $k=8$. This is why we observe non-zero rate only inside the super-critical region of $k=8$, and hence the maroon region outside of the $k=8$ super-critical region should be ignored.  Fig.~\ref{fig:surf_plots}(b) shows the optimal time-multiplexing block length $k_{(p,q)}$ for the entire $(p,q)$ space, again modulo the fact that our simulations stop at $k=8$. Between the percolation curves for two consecutive $k$ values $k_1$ and $k_2=k_1+1$, $k_{(p,q)} = k_2$, as expected. This plot can be used by the network design engineer as a tool to decide the optimal degree of time-multiplexing if $p$ and $q$ are known.

\begin{figure}[htb]
     \centering
     \includegraphics[width=0.5\textwidth]{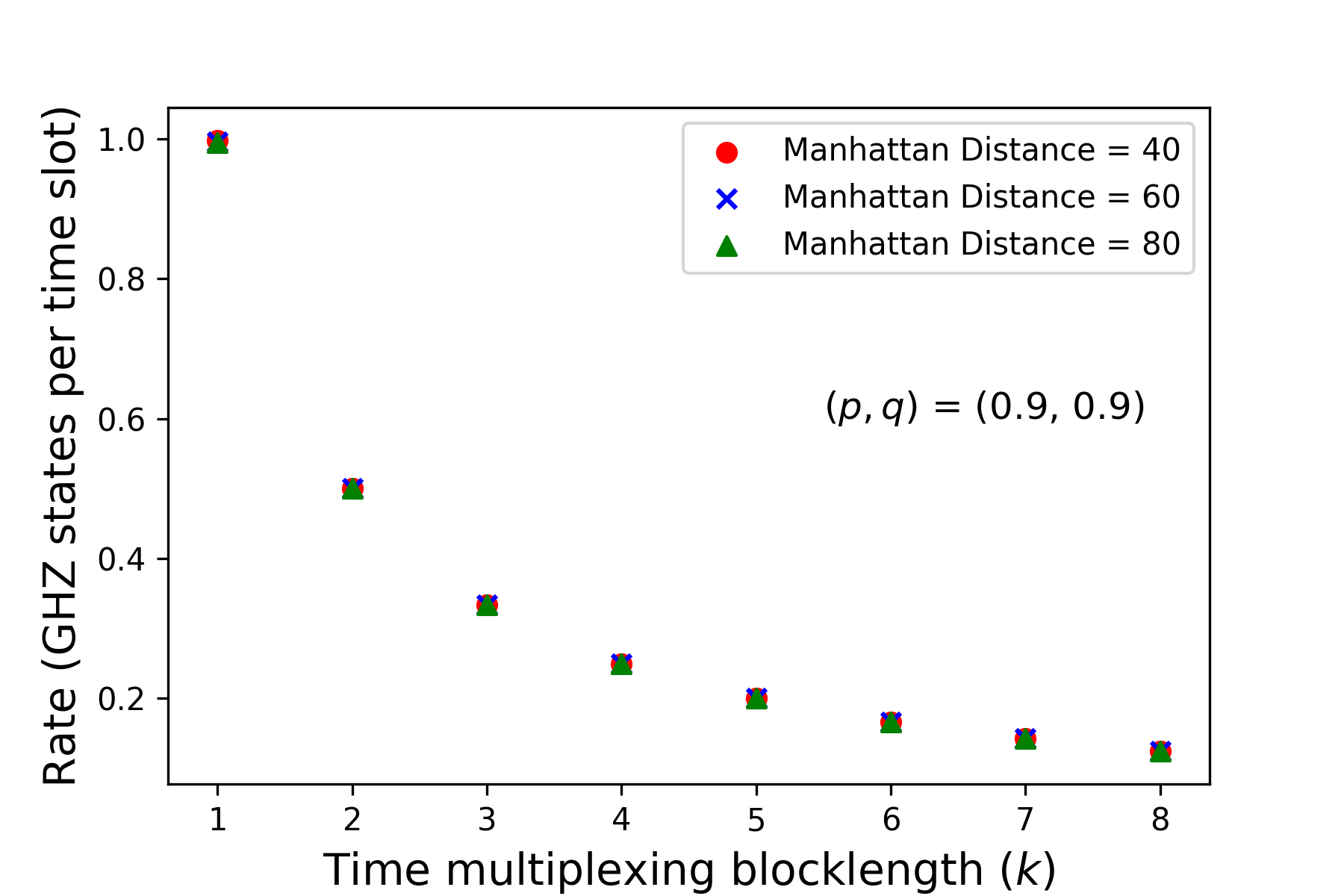} 
    \caption{Entanglement rate is plotted as a function of the time-multiplexing block length $k$ for $(p,q) = (0.9,0.9)$, plotted for three different distance between the communicating parties Alice and Bob. Since $(p,q) = (0.9,0.9)$ falls in the super-critical regions for the percolation problems for all $k \in \left\{1, \ldots, 8\right\}$, the rate is seen not to be dependent on the distance between the end users Alice and Bob.}
\label{fig:4GHZ_compare_rate_d}
\end{figure}
In Fig.~\ref{fig:4GHZ_compare_rate_d}, we plot the entanglement generation rate for three different distances between the communicating parties, to show that the rate does not change with increasing (Manhattan) distance between Alice and Bob, since $(p,q) = (0.9, 0.9)$ is within the super-critical regions of all values of $k$, the time-multiplexing block lengths, between $1$ and $8$ that are taken.

\section{Memory decoherence}
\label{sec:decoherence}
In Section~\ref{sec:time_mux}, we evaluated the performance of our $(n,k)$-GHZ protocol assuming ideal quantum memories that can hold qubits noiselessly for ${\tau_L} + (k-1)\tau$ seconds. Our analysis showed that by increasing the time-multiplexing block length $k$, we can in principle indefinitely expand the super-critical region of percolation ans hence achieve a distance-independent rate for arbitrarily low $p$ and $q$. But increasing $k$ requires higher memory coherence times as said above. In this section, we consider more realistic quantum memories that have finite coherence times, and assess how the optimal time-multiplexing block length $k_{(p,q)}$ is not only a function of $p$ and $q$, but also a function of the mean coherence time of a memory, i.e., how long a memory can (on an average) hold on to a qubit noiselessly. 
\subsection{The memory decoherence model}
\label{sub:deco_model}
We assume a step-function heralded decoherence model wherein each qubit is held perfectly in a quantum memory for a (discrete) number of time slots that is an exponentially-distributed random variable with mean $\mu$, and after that the qubit is measured off in the Pauli-X basis. This model is a good first-order approximation of common quantum memory technologies, such as color centers in diamond and trapped ions, which exhibit exponentially decaying qubit fidelities with a measurable time constant~\cite{Wang2021-vk}. The protocol stays exactly as described in Section~\ref{sub:time_mux_protocol}, wherein each repeater node only employs local link-state knowledge in forming its fusion groups, and performs $X$ measurements on unused qubits within each $k$ time slot block. The only difference between the protocol described in Section~\ref{sub:time_mux_protocol} is that there may be a few additional qubits within each $k$-time-slot block---the ones which happen to decohere---on which Pauli-X measurements are performed as well. But these decoherence-induced Pauli-X measurements are (a) performed {\em before} the fusion-groups are formed at the end of each $k$ time-slot block, and (b) a repeater node does {\em not} have information about which qubits decohered at any neighboring node. Giving the repeater node this information would add a latency of $\tau_L$ to the protocol, and more qubits could be lost to decoherence while waiting for this information to arrive, hence defeating the purpose of benefiting from time multiplexing.


Another thing to note is that our mention of `qubit' in the paper should be interpreted as being logical (error-corrected) qubits, and not physical qubits. This is because loss on a logical qubit can be heralded with the error-detection property of a quantum code. This is an important assumption in the Pauli-X measurement based model of lost qubits in the preceding paragraph. Unheralded qubit loss would end up destroying the entanglement in the entire network. Our qubits could be encoded in loss-resistant error correcting codes such as a tree code~\cite{varnava2006loss} or a crazy graph code, which can correct for unheralded photon loss as well as heralded qubit loss~\cite{rudolph2017optimistic,morley2019loss}. With an appropriately-high coding overhead (ratio of logical to physical qubits in the code), and fault-tolerant $n$-GHZ  projective measurements, it will be possible to keep the entanglement intact in the network shared by Alice and Bob, despite unheralded losses. A detailed analysis of the optimal codes and the associated coding overhead that would achieve such performance for specific memory decoherence models and gate error models, remains an open research problem.

\subsection{Results}
\begin{figure}
    \centering
    \includegraphics[scale = 0.55]{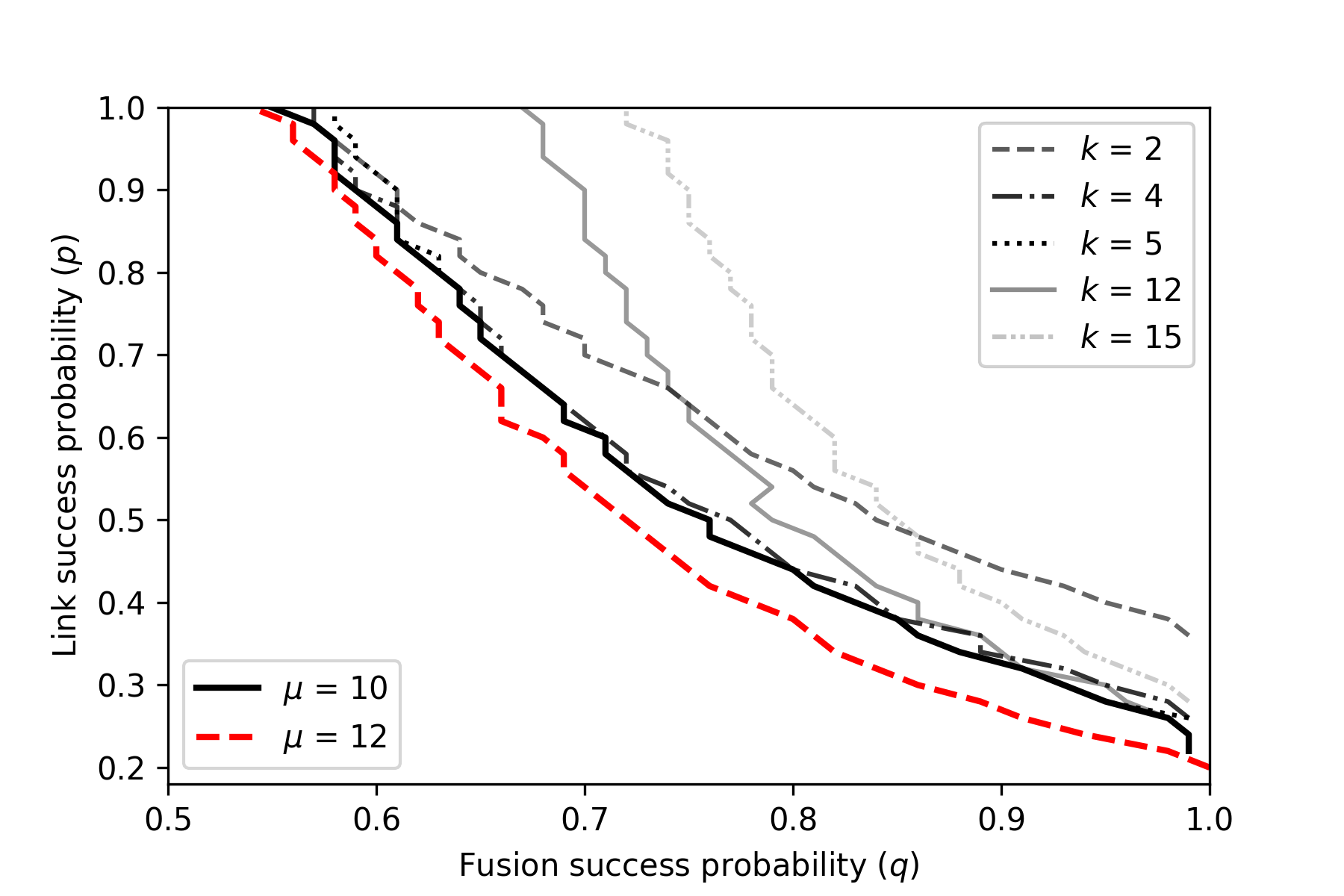}
    \caption{Site-bond percolation envelopes shown as black solid and red dashed lines when the quantum memories have finite coherence time, distributed exponentially with mean $\mu=10$ and $\mu=12$, respectively. Increasing memory coherence time improves the super-critical region, as expected. The site-bond percolation curves for  different values of $k$ when $\mu=10$ are shown as grey lines. The super-critical region expands with increasing $k$ till $k=\mu/2$ and starts shrinking if $k$ is increased further. }
    \label{fig:mem_Deco_perc}
\end{figure}

\textcolor{black}{In Fig.~\ref{fig:mem_Deco_perc}, we keep $\mu$, the mean decoherence time of each qubit constant and vary the time-multiplexing block length $k$ to plot the envelope of site-bond critical boundaries (solid black and dashed red lines) for our $(n,k)$-GHZ protocol, assuming memory decoherence as described above. Note that, $\mu/2$ is the average time length for which a link survives after being created. We observe that decreasing $\mu$ shrinks the envelope implying the network needs to have high $p$ and $q$ to achieve distance independent rate if memory coherence time is small. If $k$ is increased for a fixed $\mu$, the super-critical region expands until $k$ reaches a threshold $k^*(\mu) = \mu/2$. When $k>\mu/2$, on an average, only the links created in the previous $\mu/2$ time-steps survive the memory decoherence. If a qubit of a link is lost at a repeater, since this information is heralded and known to the repeater, it uses it in forming its fusion groups. But this information does not get communicated to the neighboring nodes. The neighboring node that holds the other qubit of the link whose one qubit is lost, it may unwittingly include that surviving qubit in one of its fusion groups, assuming the corresponding link is alive. This fusion thus does not contribute to expanding the reach of entanglement in the direction of the lost qubit. Hence, the remaining fusions in the network need to succeed with a higher probability ($q$) compared to when $k=\mu/2$ to get at least one GHZ state shared between the consumers, shrinking the super-critical region.} 

\begin{figure*}[hbt]
\centering
\subfloat[\label{sfig:fixed_k}]{%
  \includegraphics[scale = 0.6]{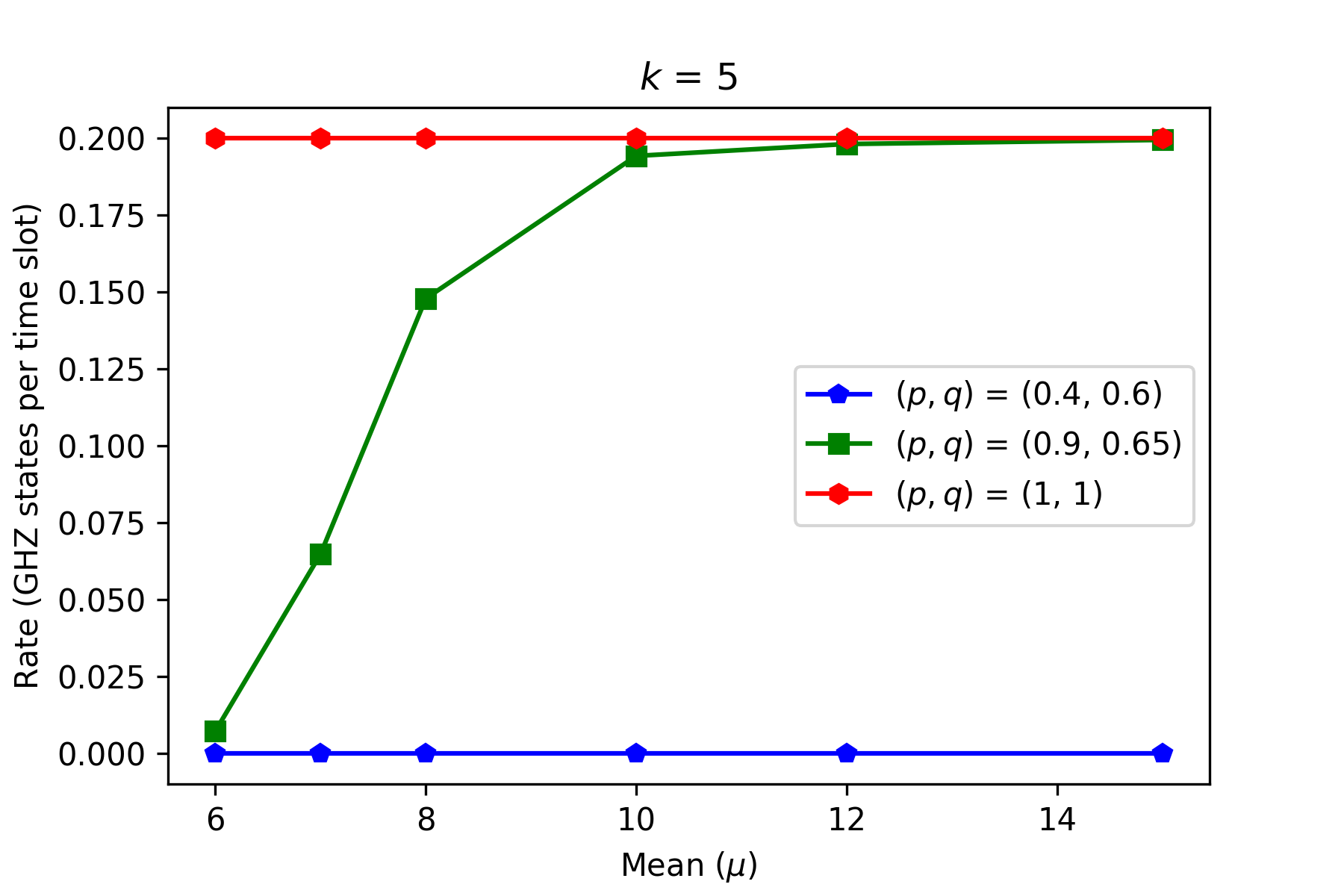}%
}
\subfloat[\label{sfig:fixed_mu}]{%
  \includegraphics[scale = 0.6]{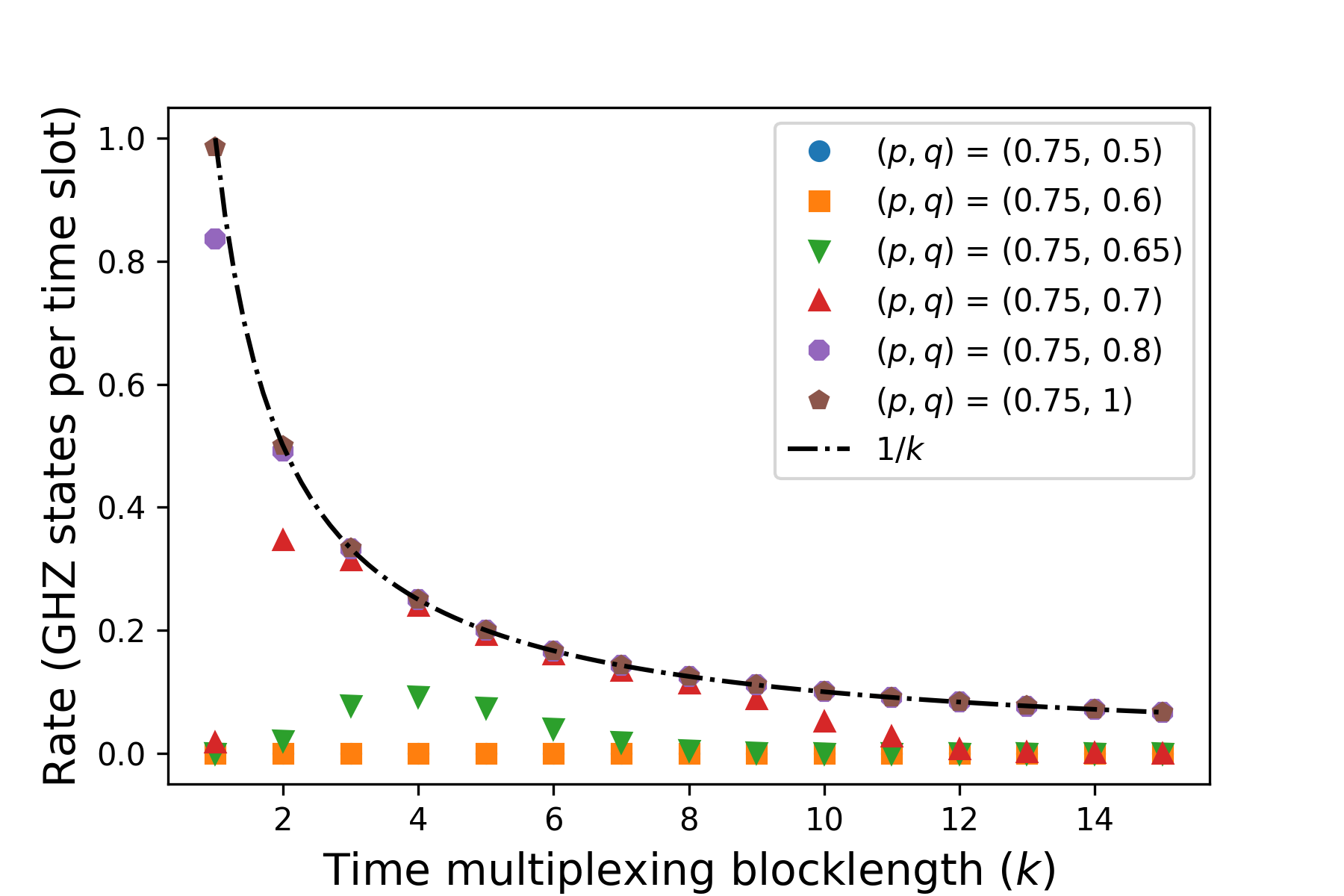}%
}
\caption{(a) Rate as a function of the average memory coherence time $\mu$ when $k=5$ for $(p,q)$ pairs in three different regions of the $(p,q)$ space (b) Rate versus $k$ for a fixed $\mu=10$. There is a $k_{(p,q,\mu)}$ associated with every$(p,q)$ and $\mu$ that achieves the best possible rate. Increasing $k$ further only hurts the rate.}
\label{fig:mem_Deco_rate}
\end{figure*}

We plot rate versus $\mu$ for three fixed $p, q$ and $k = 5$ in Fig~\ref{sfig:fixed_k}. The $(p, q)=(0.4,0.6)$ pair falls out of the super-critical regions of all $\mu$ values when $k=5$, giving exponentially decaying rate. When $(p, q)=(1,1)$, the network always percolates. As a result, the rate is always $1/k=0.2$ GHZ states per time slot. For $(p, q)=(0.9,0.65)$, the rate monotonically increases with $\mu$, and after $\mu=10$, the increase in rate shows diminishing returns. This $(p, q)$ pair is just outside or on the boundaries of the site-bond curves when $\mu<10$. It just enters the super-critical region when $\mu=10$, and stays inside, making the rate $1/k=0.2$ GHZ states per time slot for $\mu>10$. 

Fig~\ref{sfig:fixed_mu} shows rate as a function of time-multiplexing block length $k$ for a fixed $\mu=10$. If  $(p, q)$ is very deep into the super-critical region for all values of $k$, such as when $(p, q)=(0.75,1)$, the rate isn't affected by memory decoherence and follows the $1/k$ curve. If $(p, q)$ is inside the super-critical region of only a few $k$ values, the rate increases with $k$ until $k$ reaches a certain value, what we call $k_{(p,q,\mu)}$. If we keep increasing $k$ such that $k>k_{(p,q,\mu)}$, the rate starts falling, eventually reaching 0. For example, when $(p, q)=(0.75,0.65)$, the rate is maximum for $k_{(0.75,0.65,10)} = 4$.

\label{sub:deco_results}

\section{Network division}
\label{sec:net_div}
In the $n$-GHZ protocol discussed in \cite{patil2020entanglement}, which is the $(n,1)$-GHZ special-case of the time-multiplexed protocol described in Section~\ref{sub:time_mux_protocol}, the Bell pairs shared between neighboring repeaters are fused using $n$-fusions at every repeater that is not a consumer. At the end of each time slot, the consumers end up with shared GHZ state(s) with some probability. The expected number of these shared GHZ states per time slot is the entanglement rate of the protocol. This protocol translates to a site-bond percolation problem on the lattice created after performing fusions such that the site and bond occupation probabilities of the percolation problem are the fusion and link success probabilities, i.e., $q$ and $p$ respectively. The number of disjoint components in the lattice that contain both consumers after performing the fusion gives the number of GHZ states shared between the consumers. In the supercritical regime of a most standard percolation problems, such as site, bond, and site-bond percolation, there is with high probability only one connected component whose size is proportional to the number of nodes in the network. This unique connected component is known as the {\em giant connected component}. This feature of percolation limits the rate of the protocol to one GHZ state per time slot in the super-critical regime. This upper-bound on the rate is independent of network topology. However, the maximum achievable rate for a given network is given by the min-cut of the network~\cite{pirandola2019end}. For example, with $p=q=1$, the square grid network should be able to achieve a rate of $4$ GHZ states per slot. However, a randomized protocol as ours in this paper, would only achieve a rate of one GHZ state per slot, even for $p=q=1$. In this section, we show how spatial-division multiplexing can help go beyond $1$ GHZ state per slot, even without time multiplexing (i.e., for $k=1$), for non-trivial values of $p$ and $q$.

\subsection{The network-division protocol}
\begin{figure}[htb]
    \centering
    \includegraphics[scale=0.4]{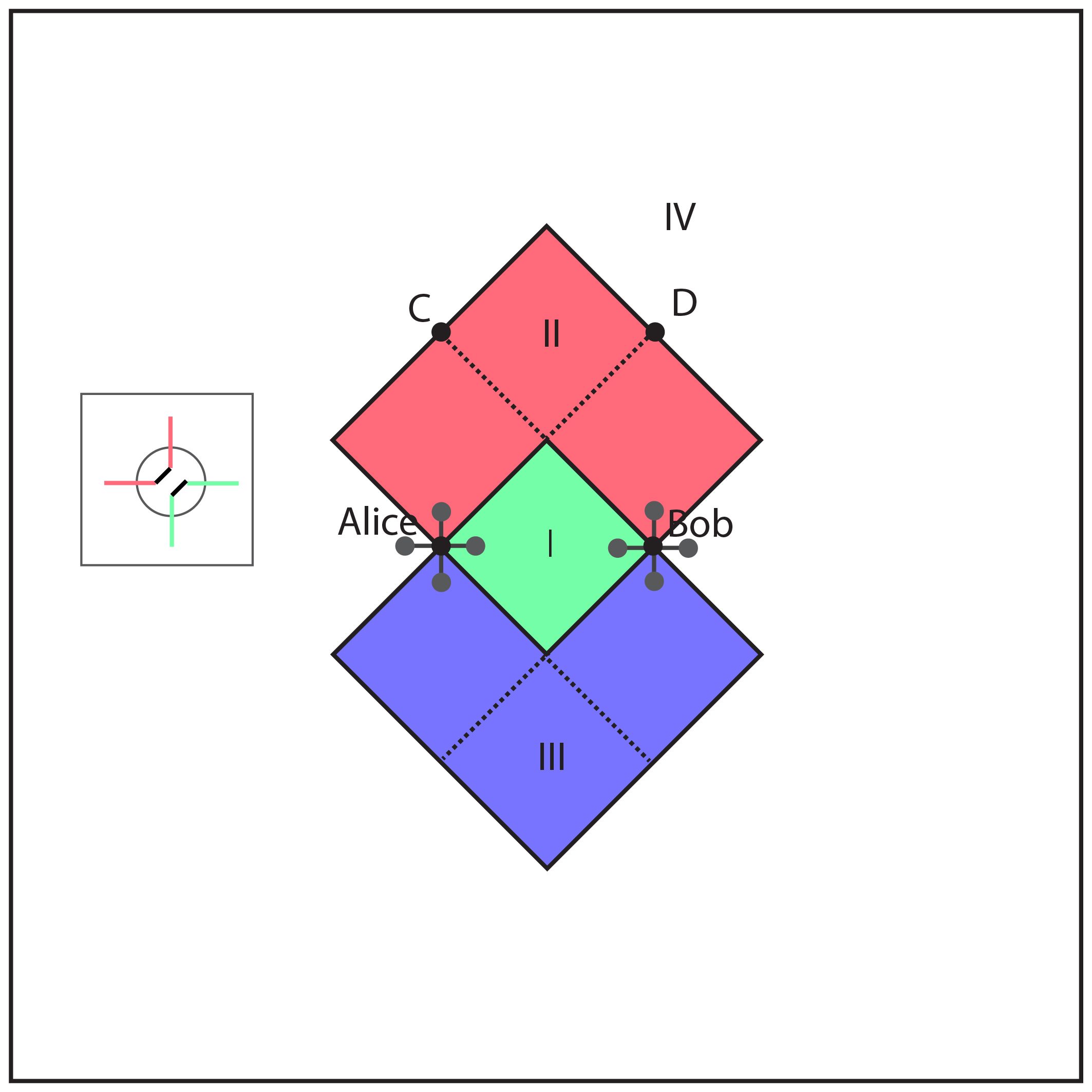}
    \caption{A square grid network divided into four regions shown in four different colors - green, red, blue and white with exactly one neighbor (grey dots) of Alice and Bob each. Repeaters on the boundaries perform fusions only between the links belonging to the same region. The inset shows a repeater on the boundary of region I and II. It disconnects the two regions by performing BSMs on the links from the same region shown in identical colors.}
    \label{fig:net_div_schm}
\end{figure}

The rate of the $n$-GHZ protocol of~\cite{patil2020entanglement} can be increased in the large $(p,q)$ region by enforcing more than one disjoint components to be created between Alice and Bob. In this section, we describe one possible method to achieve that. We divide the square-grid network into four regions as shown in Fig.~\ref{fig:net_div_schm}. This method can also be seen as spatial-division multiplexing. We define region I (shown in green) such that it is a square region and Alice and Bob are located at its diagonally opposite corners. Region II (shown in red) is composed of three copies of region I. Region III is identical to region II in shape but on the opposite side of region I. The rest of the network (the white space in Fig.~\ref{fig:net_div_schm}) constitutes region IV. The quantum repeaters on the boundaries of the four regions are not allowed to fuse links from different regions. This is equivalent to splitting every repeater on the region boundaries into two internal repeaters that act independent of one another, forming fusion groups within their own respective regions. For the square grid, each of these two internal repeaters can perform only a $2$-fusion, i.e., a BSM. This rule effectively disconnects one region from its neighboring regions and each region yields up to one shared entangled states. If a $(p,q)$ value results in more than one region to be in its respective critical region, the network would attain a distance independent rate that is at least the number of such (percolated) regions. The original $n$-GHZ protocol of ~\cite{patil2020entanglement} is performed inside each region separately. If $(p,q)$ is inside the super-critical region of region I of the protocol, i.e., the underlying lattice percolates from Alice to Bob, we can say that in region II, the lattice percolates from Alice to repeater C, then from C to D, and from D back to Bob. We can similarly divide the white region in smaller regions of the dimensions of region I and ensure percolation between Alice and Bob. We can conjecture that this construction can be extended to arbitrary locations of A and B. Note that, as Alice and Bob both can have at most one link per region each, the entangled state created in each region can only be a two-qubit state or a Bell state. This restriction in effect shrinks the percolation boundaries for each region (compared to that of the overall network), but for a high-enough $(p,q)$ opens up the possibility of multiple percolating regions and hence a rate that is higher than $1$ GHZ state per slot.

This network division rule can be easily generalized to any regular network topology such that every region of the network has exactly one link each from Alice and Bob. The maximum rate achieved for this rule is equal to the node degree of the network topology, which is also equal to the min-cut of the network. If an $n$-GHZ protocol is performed on a regular network topology with node degree $d$ such that $n<d$, we would modify the fusion rule such that the neighbors of the consumers always use the links shared with the consumers, if successful, for fusion. If the neighbors do not end up using the links with the consumers, the entanglement rate immediately drops to zero in that region. In case of random graphs such as the configuration graph, although it is always possible to divide the network into different regions, the construction of these regions is not obvious. Even for the square-grid lattice, it is very likely that our region subdivisions are not optimal. Moreover, the optimal division of the network into regions may be a function of $p$ and $q$, and will also depend upon $k$, when we employ this space-division strategy to the time-multiplexed $(n,k)$-GHZ protocol. In recent work, Manna and Ziff analyzed a new percolation problem~\cite{Manna2020-iw}, wherein $m$ points separated by a distance proportional to the lattice size $L$ simultaneously connect together, or a single point at the center of a lattice connects to the boundary through adjacent connected points of a single cluster. We believe that a variant of such a percolation problem can help us study the percolation problem induced by our space-division-multiplexed protocol within each disconnected region.
\subsection{Simulation results of entanglement rates}
\begin{figure*}[htb]
\centering
\subfloat[\label{sfig:Split_network_4GHZ_rate_p_q_L150}]{%
  \includegraphics[scale = 0.6]{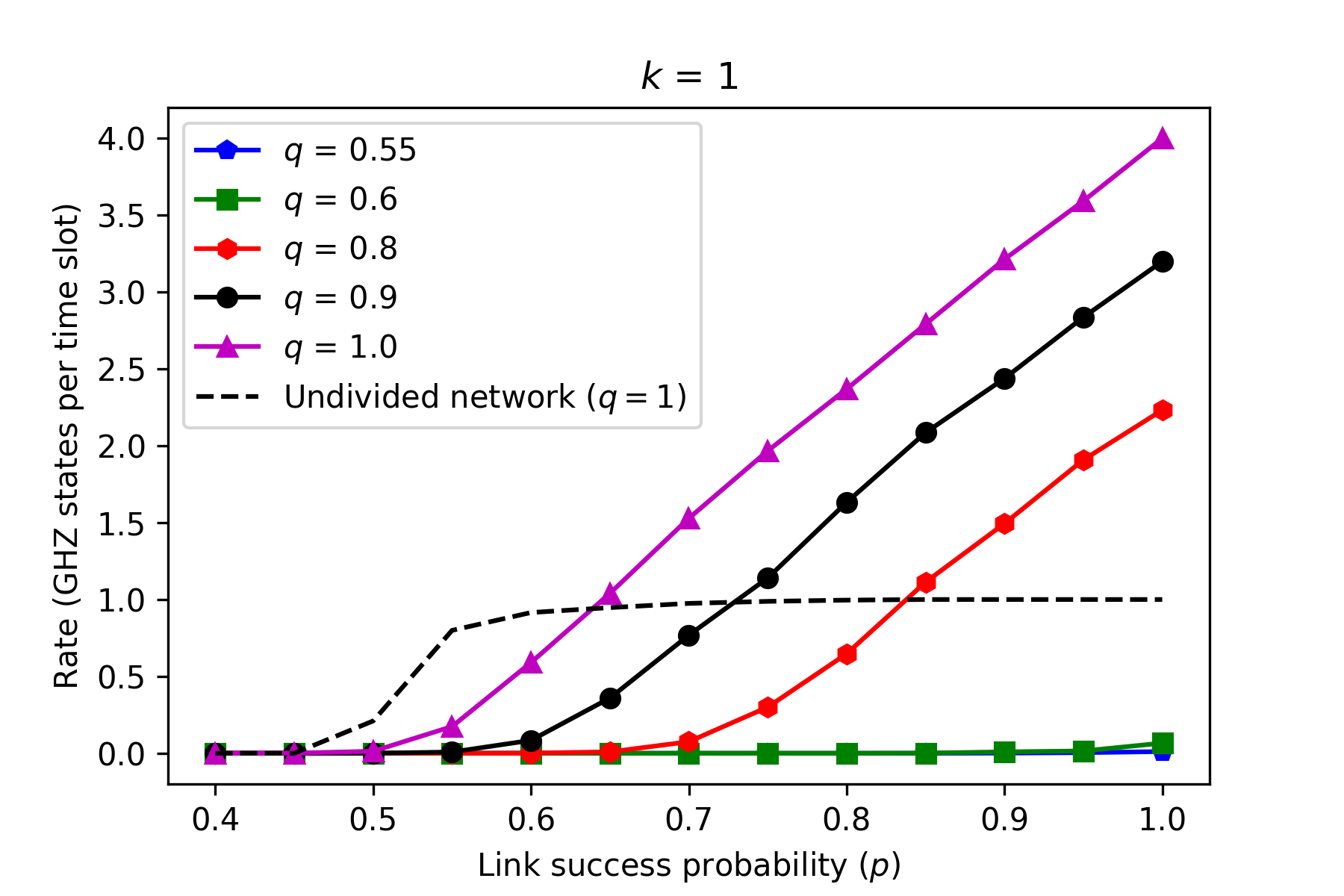}%
}
\subfloat[\label{sfig:Split_network_4GHZ_rate_q_p_L150}]{%
  \includegraphics[scale = 0.6]{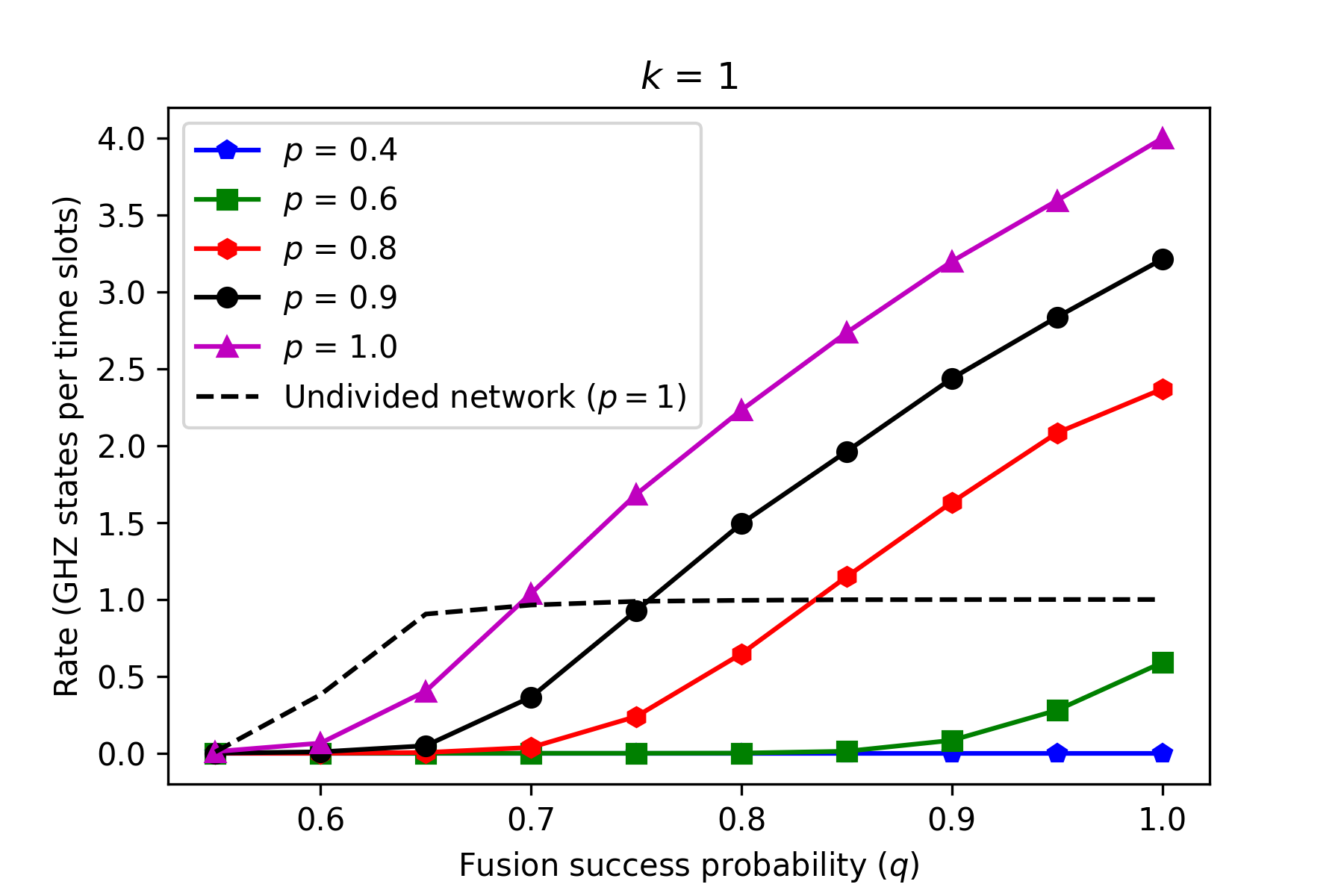}%
}
\caption{(a) Rate vs. link success probability $(p)$ and (b) Rate vs. fusion success probability $(q)$ for 4-GHZ protocol simulated on a 150$\times$150 square grid network divided into four regions as shown in Fig.~\ref{fig:net_div_schm}. Alice and Bob are located at a Manhattan distance of 25 in the same row, sufficiently inside the network to ensure that region IV can be divided into squares of region I. For some values of $(p,q)$ the undivided network achieves better rate than the divided network.}
\label{fig:net_div}
\end{figure*}
In Fig.~\ref{fig:net_div}(a), we plot the entanglement rate as a function of $p$ for fixed $q$ values, and in Fig.~\ref{fig:net_div}(b), we plot the entanglement rate as a function of $q$ for fixed $p$ values. We observe that the network division protocol achieves the rate of $4$ GHZ states per slot at $p=q=1$, the maximum rate achievable in a square grid network. The rate plots show a percolation transition, but of a different critical exponent compared to the percolation seen earlier for the $(n,k)$-GHZ protocol. Each region has its own critical boundary in the $(p,q)$ space, which we have not evaluated. The total rate is sum of rates of the four regions. When the rate is above $2$, e.g., for $(p,q) = (0.85,9)$, it implies that at least two regions are in their respective super-critical regions.

Furthermore, we observe that for $(p,q)$ values just inside the super-critical region, the rate attained by dividing the network is lower than the undivided network. Hence, for low $(p,q)$ values, spatial division multiplexing is not always advisable. This is because that $(p,q)$ which was inside the super-critical region of the undivided network may fall out of the super-critical regions of {\em all} the four regions. On the other hand, if $(p,q)$ is very high, spatial division multiplexing can afford far superior rates than the $n$-GHZ protocol. Note that time-multiplexed $n$-GHZ protocol can be run in each region to improve the super-critical site-bond region further. The exploration of this is being left for future work.


\section{Conclusions and Discussion}
\label{sec:conclusion}
It was recently shown that by allowing {\em quantum switches}---generalization of a quantum repeater that perform dynamically-switchable quantum measurements---in a quantum network to perform multi-qubit joint projective measurements in the GHZ basis, the quantum network can achieve distance-independent entanglement rate that can vastly surpass the exponentially decaying rate in a certain region of link success probability $p$ and fusion success probability $q$, called the {\em super-critical} region of a new mixed-percolation problem~\cite{patil2020entanglement}. This parameter region is determined by the percolation theory. In this paper, we showed that this super-critical region can be expanded by increasing the effective $(p,q)$ by use of time multiplexing~\cite{dhara2021sub}. As a result, the time-multiplexed protocol attains distance-independent entanglement rate for values of $(p,q)$ for which the network would have experienced exponentially decaying rate if not for the utilization of time multiplexing. We define an $(n,k)$-GHZ protocol, wherein a switch can perform a fusion measurement on up to $n$ locally-held qubits, and it chooses which groups of qubits to fuse after every subsequent block of $k$ time slots. For a given network topology, every $(p,q)$ pair has an optimal value of time multiplexing block length $k_{(p,q)}$ associated with it. Increasing the time-multiplexing block length $k$ beyond that only hurts the rate, as it starts decaying as $1/k$ GHZ states per slot. If the $(p,q)$ parameters are known, we have designed a method to calculate the degree of time-multiplexing that should be used in the network to attain the maximum possible rate for our $(n,k)$-GHZ protocol.

The rate achieved by our $(n,k)$-GHZ based time-multiplexing protocol could be improved further by restricting the repeaters to mix-and-match links generated in specific groups of time-steps. For example, if $k=8$, then the repeaters could fuse links created in the first four time slots with each other and those created in time slots 5-8 with one another. This protocol would achieve the rate that is twice that of the $(n,4)$-GHZ protocol rate. But it also comes with a trade off: if $(p,q)$ is outside the super-critical region for $k=4$, dividing the links into two groups may not be the best idea. Ideally, one could do an integer partitioning of $k$ such that there are $n_k$ partitions and if the network percolates at $(p,q)$ for the smallest partition, the rate achieved by the partitioned protocol is $n_k$ times the $(n,k)$-GHZ protocol.

While time-multiplexing helps to improve the rate in low the $(p,q)$ regime, it doesn't help much when $(p,q)$ is high. We observed that, for the $n$-GHZ protocol, i.e., the $k=1$ special case of our $(n,k)$-GHZ protocol, space-division multiplexing, i.e., restricting the flow of entanglement to certain sub-regions of the network, boosts the rate for when there is too much connectivity in the network caused due to high $(p,q)$ values. It results in multiple independent entangled states to be created between Alice and Bob. We show that by strategically dividing the network into disconnected regions, it is possible to achieve the entanglement capacity (the min-cut rate)~\cite{pirandola2019end}---the maximum achievable capacity of the network assuming ideal repeaters---at $p=q=1$. We believe that the min-cut rate of $4$ GHZ states per slot for the square grid may be achievable for $(p,q)$ strictly less than $(1,1)$ with the use of optimal time and spatial multiplexing. This will have to involve the study of percolation properties of the network sub-regions, which could potentially benefit from some recent work in percolation theory~\cite{Manna2020-iw}. We leave the exploration of this to future work.

Finally, we break the assumption of having ideal quantum memories with infinite coherence time and use realistic quantum memories whose coherence times are exponentially distributed random variables with mean $\mu$. The super-critical region expands with increasing $k$ until it reaches the average link survival time, $\mu/2$. The region shrinks when $k>\mu/2$ as the network requires the fusions to succeed with a higher probability to produce end-to-end shared entangled states. We find that if the network is deep into the super-critical region of percolation for a given $k$, i.e., inside all the percolation curves for different time-multiplexing block lengths and a fixed $\mu$ memory, decoherence doesn't hurt the rate. The rate decays as $1/k$, similar to the case when ideal memories are used. However, for other $(p,q)$ values, there is an optimal time-multiplexing block length ($k_{(p,q,\mu)}$), which is a function of $\mu, n$ and the network topology that achieves the maximum rate. The rate starts decaying if $k$ is increased beyond $k_{(p,q,\mu)}$.

Although our memory decoherence model is an improvement over using ideal memories, it is still a very simplified model. A better decoherence model would include links established with sub-unity fidelity at the outset, and realistic continuous-time noise mechanisms such as depolarization noise acting on each qubit while it is held in a quantum memory. Multi-path entanglement routing using BSMs on links with sub-unity fidelity was recently studied in~\cite{victora2020purification} by optimizing the scheduling of BSMs and Bell state purification in the network. Studying the performance of the $(n,k)$-GHZ protocol under this decoherence model would require scheduling GHZ-state purifications and $n$-fusions in the network. Refs.~\cite{GHZpurification1, GHZpurification2} study GHZ purification circuits, but they require quantum states of the same sizes to undergo purification. It may not always be possible to get two or more GHZ states of the same size shared between two repeaters. Hence, one needs to design GHZ purification circuits that can work with GHZ states of different sizes. One can go a step further and include gate imperfections, i.e., noisy gates to perform the GHZ entanglement using local operations and classical communications (LOCC), in the analysis and performance evaluation of entanglement distribution in a quantum network.

\section*{Acknowledgment}
The authors thank Gayane Vardoyan for insightful discussions.

\bibliographystyle{IEEEtran}
\bibliography{IEEEabrv,bibFile}

\appendix

\section{Percolation problem for the time multiplexed protocol -}
\label{apx:perc}
In this section, we describe the percolation problem corresponding the $(n,k)$-GHZ protocol. To generate the lattice on which we perform percolation - 
\begin{itemize}
    \item for each repeater pair, create $k$ bonds, corresponding to generation of $k$ links i.i.d with probability $p$.
    \item We treat every fusion as a site in the lattice. Hence, total number of sites are $kN$ where, $N$ is the number of repeaters in the network.
    \item Each repeater can have maximum $k$ sites corresponding to $k$ potential fusions. 
    \item In the protocol, every repeater tries to maximize the number of $n$-fusions. Hence, for a repeater pair, if $k_s$ out of $k$ links succeed, we assign these links to first $k_s$ sites at both repeaters.
    \item Each repeater randomly chooses a link from all successful links to a neighbouring repeater. Hence, we randomly pair these $k_s$ sites from both the repeaters. It means, we randomly assign neighbourhood to the $k_s$ sites. 
    \item Randomly generated neighborhood doesn't matter for sites corresponding to the failed links as these links never participate in fusions.
    \item A new lattice with new neighbourhood is created for every iteration of the protocol due to the dynamic or random way, in which links are chosen to undergo fusions.
\end{itemize}
Once the lattice is generated, we perform the usual site-bond percolation calculations corresponding to the non-time multiplexed $n$-GHZ protocol. These rules for lattice generation are applicable to both regular and random graph networks.

\end{document}